\documentclass[%reprint,
superscriptaddress,
preprint,
amsmath,amssymb,
showkeys,
hidelinks
]{revtex4-2}

\usepackage{graphicx} 
\usepackage{hyperref} 
 
\begin{document}

\title{Directional Flow of Confined Polaritons in CrSBr}

% Author: Please give full first and last names for authors and include * after the name of all corresponding authors

\author{Pratap Chandra Adak}
\email{pratapchandraadak@gmail.com}
\affiliation{Department of Physics, City College of New York, New York, NY 10031, USA}
\author{Sichao Yu}
\affiliation{Department of Physics, City College of New York, New York, NY 10031, USA}
\affiliation{Department of Physics, Graduate Center of the City University of New York (CUNY), New York, NY 10016, USA}
\author{Jaime Abad-Arredondo}
\affiliation{Departamento de Física Teórica de la Materia Condensada and Condensed Matter Physics Center (IFIMAC), Universidad Autónoma de Madrid, E28049 Madrid, Spain}
\author{Biswajit Datta}
\affiliation{Department of Physics, City College of New York, New York, NY 10031, USA}
\author{Andy Cruz}
\affiliation{Department of Physics, City College of New York, New York, NY 10031, USA}
\author{Sorah Fischer}
\affiliation{Department of Physics, City College of New York, New York, NY 10031, USA}
\author{Kseniia Mosina}
\affiliation{Department of Inorganic Chemistry, University of Chemistry and Technology Prague, Prague, Czech Republic}
\author{Zdeněk Sofer}
\affiliation{Department of Inorganic Chemistry, University of Chemistry and Technology Prague, Prague, Czech Republic}
\author{Antonio I. Fernández-Domínguez}
\affiliation{Departamento de Física Teórica de la Materia Condensada and Condensed Matter Physics Center (IFIMAC), Universidad Autónoma de Madrid, E28049 Madrid, Spain}
\author{Francisco J. Garcia-Vidal}
\email{fj.garcia@uam.es}
\affiliation{Departamento de Física Teórica de la Materia Condensada and Condensed Matter Physics Center (IFIMAC), Universidad Autónoma de Madrid, E28049 Madrid, Spain}
\author{Vinod M. Menon}
\email{vmenon@ccny.cuny.edu}
\affiliation{Department of Physics, City College of New York, New York, NY 10031, USA}
\affiliation{Department of Physics, Graduate Center of the City University of New York (CUNY), New York, NY 10016, USA}

\begin{abstract}
Nanoscale control of energy transport is a central challenge in modern photonics. Utilization of exciton-polaritons---hybrid light-matter quasiparticles---is one viable approach, but it typically demands complex device engineering to enable directional transport. Here, we demonstrate that the van der Waals magnet CrSBr offers an inherent avenue for steering polariton transport leveraging a unique combination of intrinsic optical anisotropy, high refractive index, and excitons dressed by photons. This combination enables low-loss guided modes that propagate tens of microns along the crystal $a$-axis, while simultaneously inducing strong one-dimensional confinement along the orthogonal $b$-axis. By embedding CrSBr flakes in a microcavity, we further enhance the confinement, as evidenced by energy modes that are discretized along the $b$ axis but continuous along the $a$ axis. Moreover, the magneto-exciton coupling characteristic of CrSBr allows unprecedented control over both unidirectional propagation and confinement. Our results establish CrSBr as a versatile polaritonic platform for integrated optoelectronic device applications, including energy-efficient optical modulators and switches.
\end{abstract}


\maketitle

% Text: Please use section headings and subheadings as specified below. For communications, all section headings apart from Experimental Section should be removed
% Please make the first reference to a display item bold: \textbf{Figure 1}
% Do not abbreviate Figure, Equation, etc.; display items are always singular, i.e., Figure 1 and 2.
% Equations are always singular, i.e., Equation 1 and 2, and should be inserted using the {equation} environment, not as graphics
% Please do not use footnotes in the text, additional information can be added to the Reference list.

\section{Introduction}
Exciton-polaritons, hybrid light–matter quasiparticles, offer an excellent platform for on-chip information processing, quantum photonics, and neuromorphic computing~\cite{Sanvitto2016}. 
Their hybrid character combines the high group velocity of photons, ideal for rapid signal transmission, with the interactions of excitons, necessary for nonlinear optical switching and logic. 
As bosons, polaritons can form macroscopic quantum states such as Bose-Einstein condensates and quantum superfluids, providing a basis for coherent, low-dissipation information carriers{~\cite{Deng2010}}.
By converting otherwise localized excitons with short diffusion lengths into propagating polaritons, their photonic component enables transport over tens to hundreds of microns~\cite{ ballarini2013all, ballarini2020polaritonic, sedov2025polariton}.
The realization of polaritonic integrated circuits requires platforms that are both compact and offer precise directional control of energy flow.
A common strategy to realize polaritons is to use high-finesse Fabry–Pérot cavities, with an excitonic material embedded between two distributed Bragg reflectors (DBRs)~\cite{Vahala2003a}.
While effective for confining light vertically, these structures are relatively bulky and require complex fabrication processes, limiting their integration into on-chip photonic circuits.
On the other hand, hyperbolic exciton polaritons, as recently explored in diverse materials, offer sub-diffraction confinement and high directionality~\cite{Ma2018b,Taboada-Gutierrez2020, ruta_hyperbolic_2023}.
However, the hyperbolic regime in these materials is typically narrowband, as it arises under stringent permittivity conditions, requires specialized optics, and exhibits limited propagation efficiency.

In this context, the van der Waals magnet CrSBr emerges as a unique platform that intrinsically overcomes these challenges.
As an air-stable, layered A-type antiferromagnet $(T_N\sim 132 K)$, CrSBr hosts a remarkable combination of properties ideal for guiding and manipulating polaritons~\cite{telford_layered_2020, wang_electrically-tunable_2020, lee_magnetic_2021, scheie_spin_2022, wu_quasi-1d_2022, klein_sensing_2023, wilson_interlayer_2021, bae_exciton-coupled_2022, diederich_tunable_2023, dirnberger_magneto-optics_2023, ziebel2024crsbr, datta2025magnon}.
CrSBr hosts two prominent excitons at 1.37\,eV and 1.77\,eV that persist across a wide range of thicknesses and exhibit large oscillator strengths, leading to strong light–matter interactions~\cite{wilson_interlayer_2021, dirnberger_magneto-optics_2023, datta2025magnon}.
The material’s orthorhombic crystal structure imparts pronounced in-plane anisotropy, which manifests as distinctly polarized optical absorption and photoluminescence (PL)~\cite{ziebel2024crsbr}.
This intrinsic anisotropy is inherited by the polaritons, resulting in direction-dependent dispersion and emission characteristics. 
In sufficiently thick flakes ($>100$ nm), the high refractive index of the material provides strong optical confinement, enabling the formation of self-hybridized polaritons without external cavities~\cite{dirnberger_magneto-optics_2023, wang_magnetically-dressed_2023, Li2024e, tabataba-vakili_doping-control_2024}.
Near the 1.37\,eV exciton, the opposing signs of the permittivity tensor components can even give rise to hyperbolic dispersion~\cite{ruta_hyperbolic_2023}.
Importantly, the excitonic resonances are magnetically tunable due to their coupling with the underlying spin order, allowing active control over polariton energies and dispersions via an external magnetic field~\cite{wilson_interlayer_2021, bae_exciton-coupled_2022, brennan_important_2024, datta2025magnon}.
%This field-dependence leads to a unique exciton–magnon coupling that enables optical detection of magnons and mediates novel excitonic interactions.
While these individual properties---anisotropy, strong light–matter coupling, and magnetic tunability---have been explored independently~\cite{dirnberger_magneto-optics_2023, wang_magnetically-dressed_2023, Li2024e, Li2024f}, the synergistic exploitation of CrSBr's intrinsic anisotropy and magnetic tunability to achieve and dynamically control long-range, directional polariton transport and confinement in a planar waveguide geometry remains relatively untapped.

In this work, we experimentally study the propagation and confinement of exciton-polaritons in CrSBr waveguides. 
We first investigate bare CrSBr flakes and  observe multiple polariton branches formed via strong coupling between excitons and guided modes, exhibiting highly anisotropic propagation. 
Polaritons with larger photonic components propagate tens of microns along the intermediate (\textit{a}) axis, while propagation along the easy (\textit{b}) axis remains negligible, enabling a one-dimensional confinement. 
To demonstrate additional degrees of control, we embed CrSBr flakes in a DBR-DBR cavity, resulting in quantization of the energy-momentum dispersion into discrete modes. 
Numerical simulations of the guided-mode hybridization capture both long-range polaritonic propagation in bare flakes and biaxial confinement inside the cavity. 
We further leverage the magneto-exciton coupling to tune polariton propagation and confinement using an external magnetic field.
By integrating CrSBr's intrinsic anisotropy and magnetic tunability, our work demonstrates its potential as a compact, readily-integrable, broadband platform for reconfigurable polartionic circuit applications.

\section{Results and Discussion}

\subsection{Crystalline and optical anisotropy}

CrSBr crystallizes in an orthorhombic lattice structure, with layers stacked along the \textit{c}-axis and held together by van der Waals forces~\cite{guo_chromium_2018, telford_layered_2020, wilson_interlayer_2021}.
Figure~\ref{fig:fig1}a illustrates the atomic arrangement of CrSBr in $a$--$b$ and $b$--$c$ planes.
Each individual layer exhibits structural anisotropy, with different bonding configurations along the \textit{a}- and \textit{b}-axis. 
The strong crystalline anisotropy in turn leads to markedly anisotropic optical response, which persists even away from the excitonic resonance, as evidenced by both the refractive index, $n_{aa}$, $n_{bb}$, and attenuationb constant, $\kappa_{aa}$, $\kappa_{bb}$, along the $a$ and $b$-axes, as shown in Figure \ref{fig:fig1}b. 
In the near-infrared, the refractive index along the b-axis is strongly  modified by multiple excitonic resonances, the most prominent of which is the 1.37\,eV exciton, distinguished by its large oscillator strength.
Previous studies have shown that this exciton exhibits a mixed Frenkel-Wannier character~\cite{datta2025magnon}.
Its spatial localization and strong binding energy, well below the $\sim$2\,eV bandgap, are consistent with Frenkel nature~\cite{watson2024giant, datta2025magnon}.
Meanwhile, its partial delocalization, associated with the Wannier character, renders it sensitive to long-range magnetic order. 
Consequently, a transition from antiferromagnetic to ferromagnetic order induces a 15\,meV exciton redshift and a corresponding refractive index change. 
Notably, the intrinsic crystal anisotropy also manifests in exfoliated CrSBr flakes, which adopt rectangular shapes with their longer edge aligned along the \textit{a}-axis (Figure~\ref{fig:fig1}c).

\subsection{Polariton propagation}

\begin{figure}[hbt!]
    \centering
	\includegraphics[width=16cm]{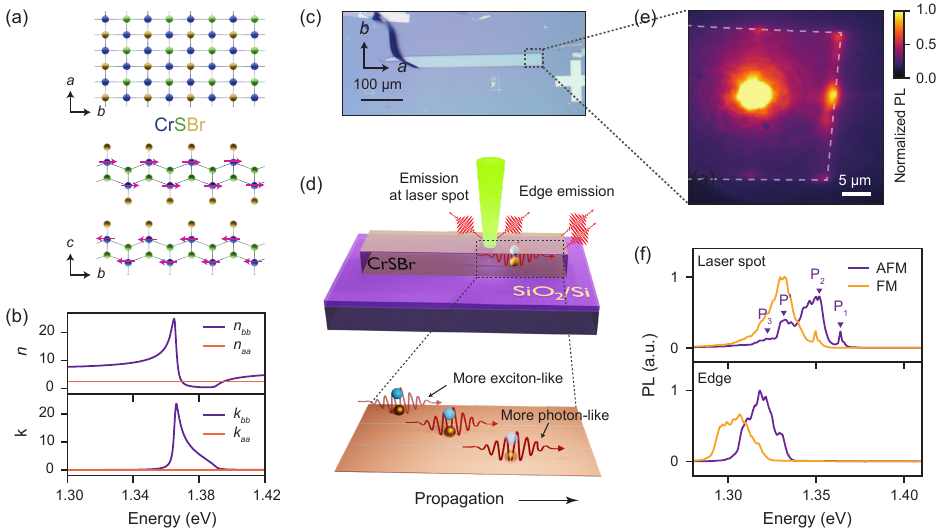}
        \caption{  Polariton propagation in CrSBr.
        (\textbf{a})~Crystal structure of CrSBr projected onto the \textit{a}--\textit{b} plane (top) and \textit{b}--\textit{c} plane (bottom). Arrows indicate the spin orientation of each layer, illustrating the AFM ordering between layers.
        (\textbf{b})~Refractive index (\(n\)) and extinction coefficient (\(k\)) of CrSBr, illustrating the optical anisotropy.
        (\textbf{c})~Microscopic image of a 90~nm thick CrSBr flake. Due to the material’s high anisotropy, the flake is oriented with its long side aligned along the \textit{a}-axis.
        (\textbf{d})~Schematic illustration of polariton propagation in the CrSBr and the measurement setup used to observe it. Upper panel shows the CrSBr flake exfoliated onto a SiO\textsubscript{2}/Si substrate and excited with a 532\,nm continuous-wave (CW) laser. Photoluminescence (PL) emission is detected both at the laser spot and at the edge of the flake due to the polariton propagation. Lower panel shows the polariton propagation with varying photonic components. More photon-like polariton branches propagate over longer distances than more exciton-like branches.
        (\textbf{e})~Real-space map of PL emission collected from the black dashed region in (c). The white dashed line marks the edge of the flake. The brightest spot at the center corresponds to emission from the laser point, while additional bright spots near the edge indicate polariton emission after propagation.
        (\textbf{f})~PL spectra collected at the laser spot (top) and the edge (bottom) for a 100~nm thick CrSBr flake in the AFM and FM state. $P_1, P_2$, and $P_3$ in the top spectrum are the peaks for the three polariton branches. $P'$ may be attributed to a defect-related state or a surface exciton.
        }
        \label{fig:fig1}
\end{figure}

To study polariton propagation in bare CrSBr flakes exfoliated on SiO$_2$/Si substrate, we measure PL from a 90~nm-thick flake excited using a 532~nm continuous-wave laser at 4~K (Figure~\ref{fig:fig1}d). 
Consistent with prior studies, we observe strong PL emission at the excitation spot, polarized along the $b$-axis, confirming the orientation of transition dipoles along the $b$-axis.
Interestingly, we also detect significant PL emission from the edges of the flake, aligned with the laser spot, along both the $a$ and $b$-axes (Figure~\ref{fig:fig1}e).
The emission tens of microns away from the excitation spot suggests that excitons couple to guided modes which propagate laterally within the flake. 
The edge emission is notably stronger along the edge parallel to the $b$-axis. This behavior directly reflects the optical anisotropy of CrSBr.
Specifically, the excitonic transition dipole moments are aligned along the $b$-axis, and therefore radiation is preferentially emitted along the orthogonal direction. This ultimately leads to greater signal collected at edges oriented along the $b$-axis.  
Around 1.37\,eV, the permittivity $\epsilon_{bb}=n_{bb}^2-\kappa_{bb}^2$ becomes negative to render the in-plane polariton dispersion hyperbolic, leading to evanescent behavior along the $a$-axis~\cite{ruta_hyperbolic_2023}.
Our far-field measurements probe a spectral regime just below this energy range, where the permittivity remains strongly anisotropic but not yet hyperbolic---resulting in highly directional, anisotropic polariton propagation.

Furthermore, measurements on multiple flakes reveal a consistent trend: PL intensity from the edge parallel to the $b$-axis not only exceeds the PL intensity at the $a$-edge, but often approaches the intensity observed at the excitation spot itself (see Figure S4, S5, Supporting Information for PL images from more flakes). 
Here we note that polaritons coupled to in-plane waveguide modes are typically inaccessible via far-field detection due to their large in-plane momentum. 
While grating couplers or near-field scanning optical microscopy (NSOM) are usually required to out-couple such modes, here the edges of the flake provide the necessary momentum mismatch via scattering, enabling detection in the far field. 
However, the efficiency of this process is inherently dependent on the specific edge geometry.
The strong edge emission suggests that the lateral guiding of polaritons is highly efficient with edge scattering providing an effective out-coupling into the far-field.
Remarkably, we observe that these exciton-polaritons can be detected tens of microns away from the excitation spot, indicating that long-range propagation is supported.

Figure~\ref{fig:fig1}f presents detailed PL spectra collected at both the excitation spot and the edge of a 100~nm thick CrSBr flake.
The spectrum at the excitation spot exhibits multiple polariton branches, arising from self-hybridization enabled by the strong refractive index contrast between CrSBr and air~\cite{dirnberger_magneto-optics_2023}.
We identify three such polariton peaks, labeled $P_1, P_2$, and $P_3$, in agreement with simulations based on the refractive index shown in Figure~\ref{fig:fig1}b. 
In addition to these peaks, we observe a weaker peak near 1.32~eV that is absent in the absorption spectra, suggesting a different origin—possibly a defect-related state or a recently reported surface exciton that does not produce a pronounced absorption dip in bulk samples (see Figure S6, Supporting Information)~\cite{shaoMagneticallyConfinedSurface2025}.
Each polariton peak appears relatively broad, likely due to a combination of strong exciton–phonon coupling~\cite{lin_strong_2024} and the presence of a broad distribution of in-plane wavevectors contributing to the emission.

Interestingly, the PL spectrum collected from the edge differs markedly, exhibiting an apparent redshift (Figure~\ref{fig:fig1}f).
Closer examination reveals that more exciton-like peaks (e.g., $P_1$) are suppressed in the edge spectrum, while the more photon-like branches (e.g., $P_3$) dominate, making it appear like a redshift.
This spatially dependent spectral redistribution reflects the fact that polariton propagation is favored for modes with higher photonic fractions—consistent with the waveguide polariton picture—and suggests a potential knob for tuning propagation via the exciton–photon admixture.
When a magnetic field is applied, both spectra undergo a redshift, a signature inherited from the excitonic component of the polaritons.
This ability to modulate polariton propagation using magnetic fields is a distinctive feature of CrSBr, made possible by its exciton-magnon coupling.

\begin{figure}[hbt!]
    \centering
	\includegraphics[width=15cm]{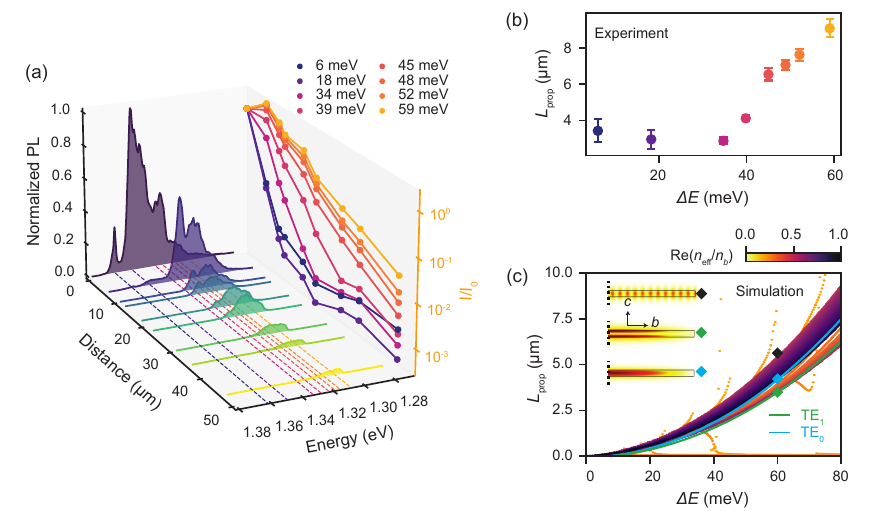}
        \caption{  Polariton propagation length.
        (\textbf{a})~ 3D plot of the PL spectra of CrSBr collected from a fixed edge (aligned along the $b$-axis) while varying the excitation spot along the $a$-axis of the flake. The horizontal axis denotes the distance between the excitation spot and the detection edge. Each spectrum is normalized to the peak intensity  at its corresponding excitation position. \textit{Inset:} Logarithmic decay of the PL intensity $I$ normalized to $I_0$, plotted as a function of energy offset $\Delta E = E_X - E$, where $E_X=1.37$~eV is the exciton energy and $E$ is the polariton energy. The intensity $I$ is taken from different spectra at various distances along a propagation path (indicated by same-colored dashed lines in the 3D plot), and $I_0$ is the intensity at the same $E$ measured at distance $= 0$. 
        (\textbf{b})~Propagation length as a function of $\Delta E$ derived from experimental data. Each data point corresponds to a decay curve of the same color shown in the inset of panel~(a).
        (\textbf{c})~Simulated propagation length as a function of $\Delta E$, showing guided modes obtained from the calculation. The color represents the effective refractive index of each mode, normalized to the refractive index along the \textit{b}-axis. The blue and green curves correspond to the dispersions of the first two $TE$ modes of a CrSBr slab waveguide.
        \textit{Inset:} Simulated electric field distributions of the $TE_0$ and $TE_1$ modes in the $b$--$c$ plane of the waveguide, along with a higher-order guided mode derived from the $TE_0$ family.
        }
        \label{fig:fig2}
\end{figure}

To further investigate the polariton propagation, we excite the sample at different positions along the $a$-axis of the flake while collecting at the same edge (parallel to the $b$-axis). 
Figure~\ref{fig:fig2}a shows the PL spectra collected at the edge of the flake as a function of the distance between the excitation and collection points.
Each spectrum is normalized with respect to the intensity maximum of the corresponding spectrum at the excitation point. 
The inset of Figure~\ref{fig:fig2}a shows the decay of the PL intensity as a function of distance from excitation to edge for different energy offset $\Delta E = E_X - E$.
Here, $E_X = 1.37$~eV is the exciton energy and $E$ is the polariton energy. 
The logarithmic scale on the $y$-axis facilitates a clearer comparison of the decay rates. 
The more exciton-like polariton branches, with small $\Delta E$, decay rapidly due to their higher component of lossy excitonic reservoirs and limited photonic character. 
In contrast, the more photon-like branches-- those with larger $\Delta E$--exhibit much longer propagation lengths and gradually dominate the emission as the distance from the excitation spot increases. 
Interestingly, for some polariton branches with large $\Delta E$, the PL intensity at a distance of 6.3~\textmu m (second point on the curve) exceeds that at the excitation spot (first point).
The vertical emission at the excitation spot is likely inefficient for these modes, while energy is efficiently channeled into lateral guided modes consistent with the waveguide-polariton picture.
%This stronger edge emission evidences the self-hybridized polaritons strongly confined along $c$-axis and coupled to guided-mode propagating laterally.

To quantify the polariton propagation, we fit the PL intensity decays with the equation, $R \propto e^{-x/L_\text{prop}}$, where $x$ is the distance from excitation to edge and $L_\text{prop}$ is the propagation length. 
Figure~\ref{fig:fig2}b summarizes the fitting results (see Section S4, Figure S7, Supporting Information for details), showing the variation of propagation lengths with $\Delta E$. 
For exciton-like branches, the propagation length remains low $\sim$2~\textmu m consistent with excitonic diffusion.
In contrast, for more photon-like branches, the propagation length increases up to 9~\textmu m.

In order to provide further insight into polariton propagation in CrSBr, we employed complementary analytical and numerical strategies (see Supporting Information for details). 
We first solved the analytical dispersion relation for transverse electric ($TE$) modes propagating along the $a$-axis in an infinite slab geometry. 
The resulting dispersion curves for the fundamental $TE_0$ and $TE_1$ modes are shown as solid lines in Figure~\ref{fig:fig2}c, demonstrating that the propagation length increases with $\Delta E$.
To connect with the experimental geometry and account for lateral confinement, we also performed numerical guided-mode simulations of CrSBr flakes using COMSOL Multiphysics. 
The corresponding propagation lengths are plotted as colored dots in Figure~\ref{fig:fig2}c shows the effective refractive index of each mode, $n_\text{eff}$, encoded in the colormap, where $n_\text{eff}$ is defined as the ratio of the propagation and the free-space wavenumbers.
For any given $\Delta E$, multiple guided modes are supported. 
These numerically determined modes can be classified as $TE_{0n}$ and $TE_{1n}$, where $n$ denotes the number of field maxima along the lateral ($b$) direction, reflecting the quantization imposed by finite flake width. 
Insets in Figure~\ref{fig:fig2}c show representative electric field distributions in the $b$--$c$ plane, illustrating that these modes clearly inherit the vertical (along $c$-axis) confinement structure of the corresponding $TE_0$ and $TE_1$ slab modes.
Although the simulated propagation lengths follow the trend of the analytical dispersion, they span a wide range, demonstrating the dependence on specific guided modes and the crucial role of flake geometry in shaping polariton transport. 
Considering the large number of supported modes, the experimentally observed propagation length reflects a weighted average over all accessible modes at a given energy. 
Overall, the simulated results show good agreement with the experiment, validating our understanding of waveguide mode dispersion in CrSBr.

\subsection{Polariton confinement}

\begin{figure}[hbt!]
    \centering
	\includegraphics[width=17cm]{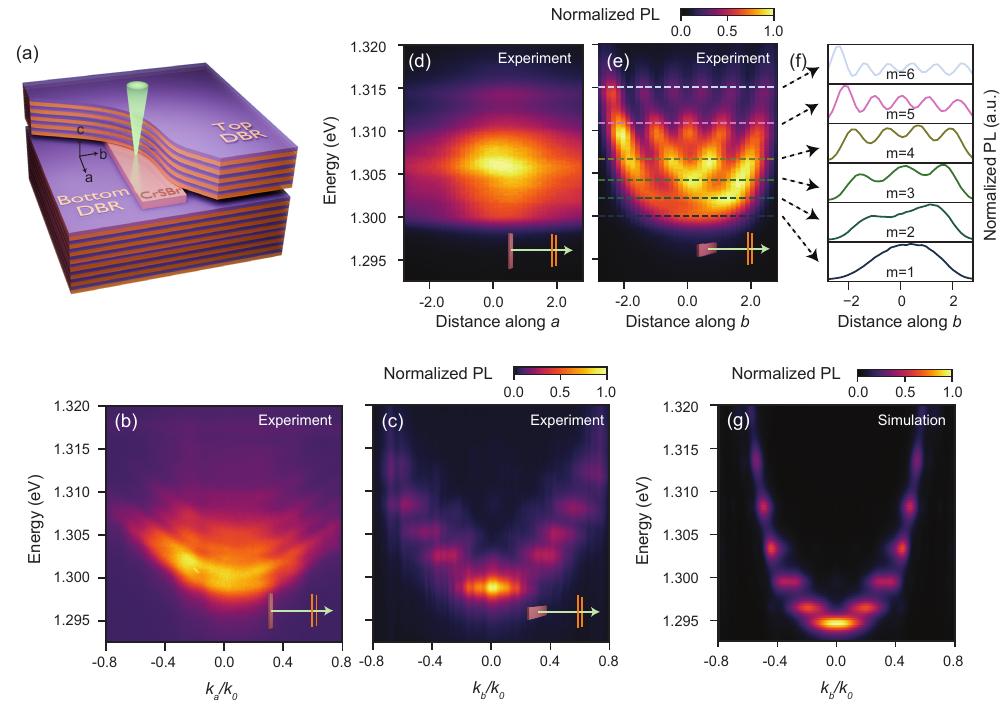}
        \caption{  Exciton-polariton confinement in CrSBr cavities.
        (\textbf{a})~Schematic illustration of the DBR/CrSBr/DBR microcavity, consisting of $\sim$130\,nm-thick, 5.0\,\textmu m-width CrSBr flakes encapsulated between two DBR mirrors.
        (\textbf{b,c})~Experimentally measured momentum-resolved PL spectra of the 5.0\,\textmu m-width CrSBr flake with momentum \( k \) aligned along the \( k_a \) (b) and \( k_b \) (c) directions in Fourier space. Here, \( k_0 = {2\pi}/{\lambda_0}\), \({\lambda_0}\) is the wavelength of the uncoupled cavity mode at \( k_{a,b} = 0\).
        (\textbf{d,e})~Experimentally measured spatially resolved PL spectrum along the \textit{a}-axis (d) and the \textit{b}-axis (e) of  the same 5.0\,\textmu m-wide flake. 
        (\textbf{f})~Spatial distribution of the measured PL intensity along the \textit{b}-axis of the CrSBr flake at energies of 1.301\,eV, 1.303\,eV, 1.306\,eV, 1.309\,eV, 1.314\,eV, and 1.319\,eV, corresponding to mode number ($m$) from 1 to 6. Each curve corresponds to the dashed lines of the same color in panel~(e).
        (\textbf{g})~Simulated momentum-resolved PL spectrum of the DBR/CrSBr/DBR microcavity with a 5.0\,\textmu m-wide CrSBr flake. Momentum \( k \) is aligned along the \( k_b \) direction in Fourier space.
        PL intensities are normalized to the maximum intensity within each respective panel.
        Insets in panels (a-d) indicate the orientation of the flake with respect to the spectrometer slit.
        }
        \label{fig:fig3}
\end{figure}

While excitons strongly coupled to propagating waveguide modes enable long-range polariton transport, we can gain further control by enhancing the confinement using external cavity.
To realize such confinement, we exfoliate CrSBr flakes onto a commercially available distributed Bragg reflector (DBR) with a stop-band centered at 940~nm, and subsequently deposit a top DBR with a stop-band centered at 930~nm.
Figure~\ref{fig:fig3}a schematically illustrates the resulting microcavity structure (see Figure S3, Supporting Information for optical microscope images of the studied flakes).
The naturally limited lateral dimension of exfoliated CrSBr flakes along the $b$-axis already provides a degree of in-plane optical confinement.
The addition of DBR mirrors on top and bottom sides of the flake enhances this effect by increasing the reflectivity at the flake boundaries, strengthening the vertical photon confinement through Fabry–Pérot resonances.

To study polariton confinement, we employ Fourier-space spectroscopy by imaging the back focal plane of the objective (i.e., the Fourier plane) onto the spectrometer CCD. 
The entrance slit is aligned along one in-plane momentum axis in the Fourier plane, while the diffraction grating provides energy resolution.
Figure~\ref{fig:fig3}b, c show the momentum-resolved PL spectra collected along the $a$ and $b$-axes, respectively, from a CrSBr flake with dimensions of 5.0\,\textmu m width, 108\,\textmu m length, and 130~nm thickness. 
Along the $a$-axis (Figure~\ref{fig:fig3}b), the exciton-polariton dispersion exhibits a continuous variation of energy with in-plane momentum, forming the expected parabolic dispersion of a planar microcavity.
In contrast, when the slit is oriented along the $b$-axis (Figure~\ref{fig:fig3}c), the dispersion becomes discretized---a hallmark of lateral confinement.
Instead of a continuous parabolic dispersion, we observe multiple discrete polariton sub-branches corresponding to quantized wavevectors $k_b$ imposed by the waveguide confinement along the $b$-axis.
The lowest energy state at $k_b=0$ is set by the cutoff energy of the cavity mode, determined by vertical confinement along the $c$-axis.
Notably, beyond the dominant parabola corresponding to the fundamental confined mode, Figure~\ref{fig:fig3}b shows additional weaker, parallel dispersion branches. 
These weaker $E$--$k_a$ dispersions correspond to modes with different $k_b$, appearing at the energies of the dominant discrete modes (see Figure S9, Supplementary Information, for enhanced visualization from a different spot). 
The presence of a continuous energy-momentum dispersion along $k_a$, and discrete confined modes along $k_b$ are further evidenced by the momentum-resolved absorption measurements (Figure S12, Supporting Information). 
Together, these results establish that the observed polariton modes arise from simultaneous confinement along the $b$- and $c$-axes, while remaining free to propagate along the $a$-axis.

The strong in-plane confinement and associated energy quantization are further corroborated by real-space measurements.
Figure~\ref{fig:fig3}d,e present color-scale plots of the normalized PL intensity as a function of energy and position along the $a$ and $b$-axis of the flake, respectively.
Along the $a$-axis, PL intensity gradually decreases away from the excitation center. 
In contrast, along the $b$-axis, PL emission is quantized, corresponding to the discrete $k_b$ values identified in the momentum-resolved spectrum of Figure~\ref{fig:fig3}c.
Strikingly, the PL intensity exhibits spatial maxima at well-defined positions along the $b$-axis, with each peak corresponding to a standing-wave mode associated with a particular quantized $k_b$.
The number of intensity nodes increases with the mode index, consistent with the expected spatial profiles of confined waveguide modes.
To clearly visualize this spatial mode structure, we plot the normalized PL intensity as a function of position along the $b$-axis at six representative energy values in Figure~\ref{fig:fig3}f.
This real-space modulation provides direct evidence of polariton confinement along the $b$-axis and confirms the formation of quantized polariton eigenmodes within the cavity structure.
These confined modes arise from an effective potential well along the \textit{b}-axis, created by the combined effects of the narrow length scale along the \textit{b}-axis and the high refractive index contrast at the interfaces.

To gain further insight into polariton confinement in CrSBr, we performed frequency-domain electromagnetic simulations using COMSOL Multiphysics (see Section S6, Figure S8, Supporting Information for details).
Figure~\ref{fig:fig3}g displays the momentum-resolved simulated PL emission from a CrSBr flake with geometry similar to the one used experimentally, encapsulated between two DBRs.
Energy–momentum dispersion is quantized, clearly revealing discrete polariton modes arising from lateral and vertical confinement.
The resonance frequency in the numerical simulations is offset by 5\,meV relative to the experimental measurements, likely due to the use of nominal flake thicknesses in the model.
Overall, the simulated dispersion agrees well with our experimental observations.

\begin{figure}[hbt!]
    \centering
	\includegraphics[width=16cm]{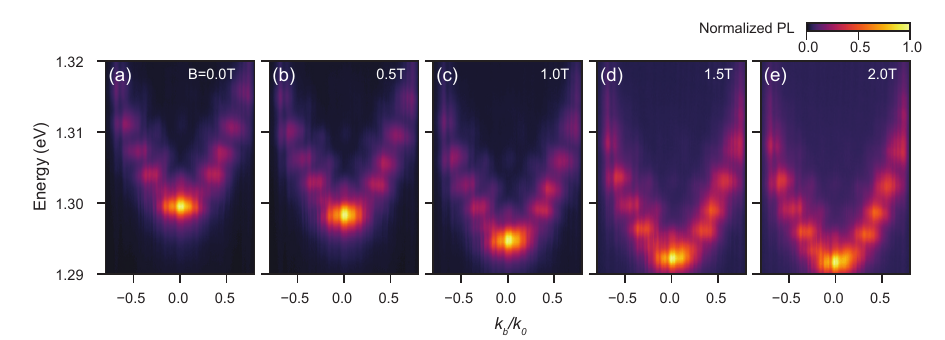}
        \caption{ Tuning polariton confinement using magnetic field.
        (\textbf{a-e})~Momentum-resolved PL spectra along the $b$-axis at magnetic fields of 0\,T (a), 0.5\,T (b), 1.0\,T (c), 1.5\,T (d), and 2.0\,T (e), applied along the \textit{c}-axis. 
        The exciton-polariton energy exhibits redshifts due to spin-canting transitions induced by the magnetic field.
        PL intensities are normalized to the maximum intensity within each respective panel.
        }
        \label{fig:fig4}
\end{figure}

By varying the width and thickness of CrSBr flakes, we can tune the confinement potential and observe corresponding changes in the discrete polariton energy spectra.
See Figure S10, and corresponding Figure S8 Supporting Information for momentum- and spatially-resolved PL measurements and simulations from another flake with width of 2.6~\textmu m, showing quantized dispersions distinct from the 5.0~\textmu m wide flake studied in Figure~\ref{fig:fig3}.
However, a more dramatic and distinctive tuning knob arises from the unique magnetic control of excitons in CrSBr.
In Figure~\ref{fig:fig4}, we present momentum-resolved PL spectra for five different magnetic fields applied along the $c$-axis (see Figure~S11, Supporting Information for corresponding real space spectra).
As the magnetic field increases, the exciton resonance undergoes a redshift (see Figure S13, Supporting Information for change in refractive index due to magnetic phase transition). 
Consequently, the confined polariton energies shift accordingly due to their hybrid excitonic-photonic nature.
We observe a clear, monotonic redshift of all quantized polariton modes with increasing magnetic field, confirming that the confined modes possess substantial excitonic character---an essential hallmark of their polaritonic nature.
These results further consolidate that, in addition to structural confinement, magnetic tuning offers a powerful external control over discrete polariton spectra in CrSBr microcavities.

These findings demonstrate the unique capabilities of CrSBr for achieving simultaneous polariton propagation and confinement with in-situ tunability. Our observation invites a broader comparison with other material platforms showing long range polariton transport. 
Long-range polariton propagation has been demonstrated across diverse systems, including organic semiconductors~\cite{rozenman2018long, hou2020ultralong, balasubrahmaniyam2023enhanced}, halide perovskites~\cite{xu2023ultrafast}, and GaAs-based microstructures~\cite{steger2013long, zaitsev2015diffusive}.
More recently, two-dimensional (2D) materials, particularly transition metal dichalcogenides (TMDs), have garnered significant attention due to their large exciton binding energies and highly tunable properties~\cite{wurdack2021motional, cho2023ultra}.
However, in addition to hosting strongly bound excitons in a van der Waals platform, CrSBr offers several distinct advantages. 
It supports excitons at any thickness, allowing thickness to be used as a design parameter to tune the polariton wavelength.
The inherent optical anisotropy, as demonstrated in our study, further imparts directional control on the guided polaritons without requiring additional device structuring. 
Most remarkably, the magnetic field offers an in-situ tuning knob on these polaritons.

\section{Conclusion and outlook}
In conclusion, we have demonstrated that the crystal anisotropy present in the van der Waals magnet CrSBr offers a built-in mechanism for engineering polariton dynamics, enabling robust propagation along a single crystal axis while simultaneously allowing for tight confinement along the other.
The 1D confinement channels are ideal testbeds for exploring polariton-polariton interactions and Bose-Einstein condensation~\cite{Zhang2025}.
Furthermore, combining this intrinsic anisotropy with magnetic control offers a compelling route towards non-reciprocal polariton devices, including optical isolators and circulators, where the underlying magnetic order breaks time-reversal symmetry. 
Overall, our work positions anisotropic magnetic semiconductors as a fertile ground for developing compact, energy-efficient, and reconfigurable photonic circuits.
% Experimental section

\section{Experimental Section}

\subsection{Fabrication details}

Bulk CrSBr crystals were obtained through a chemical vapor transport process. The CrSBr flakes for measurement were derived from exfoliation of bulk crystals. This process involved initially cleaving the crystals on scotch tape, ensuring a known and fixed crystal orientation. Subsequently, the crystals were mechanically exfoliated onto polydimethylsiloxane (PDMS). Following this step, a substrate chip (SiO$_2$/Si or DBR) was used to transfer CrSBr from PDMS to the chip. Atomic force microscopy was used to determine the thickness of the CrSBr flakes.

For CrSBr microcavity used in confinement measurements, CrSBr flakes were exfoliated on a DBR substrate which is made of 8 pairs of SiO$_2$ (159.7 nm) and TiO$_2$ (101.3 nm) layers deposited on a Si substrate (from Spectrum Thin Films, Inc). Then a top DBR with 5 pairs of SiO$_2$ (158.0 nm) and Si$_3$N$_4$ (113.9 nm) was deposited on the sample using Plasma Enhanced Chemical Vapor Deposition (PECVD) to complete the microcavity.

\subsection{Polariton propagation measurements}

Polariton propagation measurements were performed using a closed-cycle cryostat (OptiCool) at a base temperature of 2 K. A continuous-wave (CW) diode laser with a wavelength of 532\,nm was focused on the sample using a \(\times\)100 objective and used to excite it. The PL signals from the excitation point and the edge of the sample were collected by the same objective and then analyzed using a Princeton Instruments spectrometer
(Model SpectraPro HRS-500) equipped with a grating of 300 lines per millimeter (Figure S1, Supporting Information). The slit of the spectrometer was removed to ensure that all PL signals from the sample could be collected simultaneously. The spectra from the excitation point and the edge were selected manually from the raw data and analyzed using Python code.

\subsection{Spatially resolved and momentum-resolved photoluminescence spectroscopy}
The spatially resolved and momentum-resolved PL measurements for the cavity sample were conducted using a Montana cryostat at a base temperature of 4~K (Figure S2, Supporting Information). PL signals were collected by a \(\times\)100 objective and analyzed using a Princeton Instruments spectrometer (Model SpectraPro SP-2150). A k-space lens was placed in the output optical path to enable switching between spatially resolved and momentum-resolved measurements. A slit was placed in front of the spectrometer to ensure that PL signals were collected only along a specific direction of the flake. A Dove prism was used to rotate the signal beam, allowing the orientation between the flake and the slit to be adjusted.

\subsection{Magnetic field-induced shifts}
Magnetic field-induced shifts were measured in the OptiCool system at a baser temperature of 2 K. The magnetic field was applied along the hard (\textit{c}) axis of the CrSBr flake, ranging from $-2.5$\,T to 2.5\,T.

\medskip
\textbf{Supporting Information} \par %Please delete the Suppporting Information statement if it is not applicable. Please supply Supporting Information in another file. Supporting information should not be provided in .tex format
Supporting Information is available from the Wiley Online Library or from the author.

% Acknowledgements
\medskip
\textbf{Acknowledgements} \par %delete if not applicable))
P.C.A. was supported by the Army Research Office grant W911NF-23-1-0394, S.Y. was supported by the NSF grant 2216838. 
B.D. was supported by the Gordon and Betty Moore Foundation (Grant No. 12764). A. C. and V.M.M. were supported by the Department of Energy's Office of Basic Energy Sciences award: DE-SC0025302.
Z.S. was supported by project LUAUS25268 from Ministry of Education Youth and Sports (MEYS), ERC-CZ program (project LL2101) from Ministry of Education Youth and Sports (MEYS) and by the project Advanced Functional Nanorobots (reg. No. CZ.02.1.01/0.0/0.0/15\_003/0000444 financed by the EFRR). K.M. were supported from the grant of Specific university research – grant No A1\_FCHT\_2025\_013.
J.A.A., A.I.F.D., and F.J.G.V. were supported by MICIU/AEI/10.13039/501100011033 and FEDER, EU under grants PID2021-126964OB-I00 and PID2021-125894NB-I00 and by the (MAD2D-CM)-UAM7 project funded by Comunidad de Madrid, by the Recovery, Transformation and Resilience Plan, and by NextGenerationEU from the European Union. A.I.F.D. also acknowledges financial support from the European Union’s Horizon Europe Research and Innovation Programme through agreement 101070700 (MIRAQLS).

\medskip
\textbf{Author Contributions} \par %delete if not applicable))
P.C.A., S.Y., J.A-A., and B.D. contributed equally to this work.

\medskip
\textbf{Conflict of Interest} \par
The authors declare no competing interests.

\medskip
\textbf{Data Availability Statement} \par
The data that support the findings of this study are available from the corresponding author upon reasonable request.

% References
\medskip

% Use the following code if you wish to generate your bibliography with BibTeX;
% replace the string "MSP-template" below with the name(s) of
% the BibTeX data base(s) you want to use.
% The resulting bibliography-output (the content of the .bbl file)
% must be pasted back into this file before submission.
% Please also include your BibTeX data base file(s) in your submission
% so that we can re-run BibTeX if necessary.
%
\bibliographystyle{apsrev4-2}
\bibliography{references}

\begin{thebibliography}{1}
\providecommand{\url}[1]{\texttt{#1}}
\providecommand{\urlprefix}{URL }

\bibitem{Delves1967}
L.~M. Delves, J.~N. Lyness,
\newblock \emph{Mathematics of Computation} \textbf{1967}, \emph{21} 543.

\end{thebibliography}


%apsrev4-2.bst 2019-01-14 (MD) hand-edited version of apsrev4-1.bst
%Control: key (0)
%Control: author (72) initials jnrlst
%Control: editor formatted (1) identically to author
%Control: production of article title (-1) disabled
%Control: page (0) single
%Control: year (1) truncated
%Control: production of eprint (0) enabled
\begin{thebibliography}{39}%
\makeatletter
\providecommand \@ifxundefined [1]{%
 \@ifx{#1\undefined}
}%
\providecommand \@ifnum [1]{%
 \ifnum #1\expandafter \@firstoftwo
 \else \expandafter \@secondoftwo
 \fi
}%
\providecommand \@ifx [1]{%
 \ifx #1\expandafter \@firstoftwo
 \else \expandafter \@secondoftwo
 \fi
}%
\providecommand \natexlab [1]{#1}%
\providecommand \enquote  [1]{``#1''}%
\providecommand \bibnamefont  [1]{#1}%
\providecommand \bibfnamefont [1]{#1}%
\providecommand \citenamefont [1]{#1}%
\providecommand \href@noop [0]{\@secondoftwo}%
\providecommand \href [0]{\begingroup \@sanitize@url \@href}%
\providecommand \@href[1]{\@@startlink{#1}\@@href}%
\providecommand \@@href[1]{\endgroup#1\@@endlink}%
\providecommand \@sanitize@url [0]{\catcode `\\12\catcode `\$12\catcode `\&12\catcode `\#12\catcode `\^12\catcode `\_12\catcode `\%12\relax}%
\providecommand \@@startlink[1]{}%
\providecommand \@@endlink[0]{}%
\providecommand \url  [0]{\begingroup\@sanitize@url \@url }%
\providecommand \@url [1]{\endgroup\@href {#1}{\urlprefix }}%
\providecommand \urlprefix  [0]{URL }%
\providecommand \Eprint [0]{\href }%
\providecommand \doibase [0]{https://doi.org/}%
\providecommand \selectlanguage [0]{\@gobble}%
\providecommand \bibinfo  [0]{\@secondoftwo}%
\providecommand \bibfield  [0]{\@secondoftwo}%
\providecommand \translation [1]{[#1]}%
\providecommand \BibitemOpen [0]{}%
\providecommand \bibitemStop [0]{}%
\providecommand \bibitemNoStop [0]{.\EOS\space}%
\providecommand \EOS [0]{\spacefactor3000\relax}%
\providecommand \BibitemShut  [1]{\csname bibitem#1\endcsname}%
\let\auto@bib@innerbib\@empty
%</preamble>
\bibitem [{\citenamefont {Sanvitto}\ and\ \citenamefont {{K{\'e}na-Cohen}}(2016)}]{Sanvitto2016}%
  \BibitemOpen
  \bibfield  {author} {\bibinfo {author} {\bibfnamefont {D.}~\bibnamefont {Sanvitto}}\ and\ \bibinfo {author} {\bibfnamefont {S.}~\bibnamefont {{K{\'e}na-Cohen}}},\ }\href {https://doi.org/10.1038/nmat4668} {\bibfield  {journal} {\bibinfo  {journal} {Nature Materials}\ }\textbf {\bibinfo {volume} {15}},\ \bibinfo {pages} {1061} (\bibinfo {year} {2016})}\BibitemShut {NoStop}%
\bibitem [{\citenamefont {Deng}\ \emph {et~al.}(2010)\citenamefont {Deng}, \citenamefont {Haug},\ and\ \citenamefont {Yamamoto}}]{Deng2010}%
  \BibitemOpen
  \bibfield  {author} {\bibinfo {author} {\bibfnamefont {H.}~\bibnamefont {Deng}}, \bibinfo {author} {\bibfnamefont {H.}~\bibnamefont {Haug}},\ and\ \bibinfo {author} {\bibfnamefont {Y.}~\bibnamefont {Yamamoto}},\ }\href {https://doi.org/10.1103/RevModPhys.82.1489} {\bibfield  {journal} {\bibinfo  {journal} {Reviews of Modern Physics}\ }\textbf {\bibinfo {volume} {82}},\ \bibinfo {pages} {1489} (\bibinfo {year} {2010})}\BibitemShut {NoStop}%
\bibitem [{\citenamefont {Ballarini}\ \emph {et~al.}(2013)\citenamefont {Ballarini}, \citenamefont {De~Giorgi}, \citenamefont {Cancellieri}, \citenamefont {Houdr{\'e}}, \citenamefont {Giacobino}, \citenamefont {Cingolani}, \citenamefont {Bramati}, \citenamefont {Gigli},\ and\ \citenamefont {Sanvitto}}]{ballarini2013all}%
  \BibitemOpen
  \bibfield  {author} {\bibinfo {author} {\bibfnamefont {D.}~\bibnamefont {Ballarini}}, \bibinfo {author} {\bibfnamefont {M.}~\bibnamefont {De~Giorgi}}, \bibinfo {author} {\bibfnamefont {E.}~\bibnamefont {Cancellieri}}, \bibinfo {author} {\bibfnamefont {R.}~\bibnamefont {Houdr{\'e}}}, \bibinfo {author} {\bibfnamefont {E.}~\bibnamefont {Giacobino}}, \bibinfo {author} {\bibfnamefont {R.}~\bibnamefont {Cingolani}}, \bibinfo {author} {\bibfnamefont {A.}~\bibnamefont {Bramati}}, \bibinfo {author} {\bibfnamefont {G.}~\bibnamefont {Gigli}},\ and\ \bibinfo {author} {\bibfnamefont {D.}~\bibnamefont {Sanvitto}},\ }\href@noop {} {\bibfield  {journal} {\bibinfo  {journal} {Nature communications}\ }\textbf {\bibinfo {volume} {4}},\ \bibinfo {pages} {1778} (\bibinfo {year} {2013})}\BibitemShut {NoStop}%
\bibitem [{\citenamefont {Ballarini}\ \emph {et~al.}(2020)\citenamefont {Ballarini}, \citenamefont {Gianfrate}, \citenamefont {Panico}, \citenamefont {Opala}, \citenamefont {Ghosh}, \citenamefont {Dominici}, \citenamefont {Ardizzone}, \citenamefont {De~Giorgi}, \citenamefont {Lerario}, \citenamefont {Gigli} \emph {et~al.}}]{ballarini2020polaritonic}%
  \BibitemOpen
  \bibfield  {author} {\bibinfo {author} {\bibfnamefont {D.}~\bibnamefont {Ballarini}}, \bibinfo {author} {\bibfnamefont {A.}~\bibnamefont {Gianfrate}}, \bibinfo {author} {\bibfnamefont {R.}~\bibnamefont {Panico}}, \bibinfo {author} {\bibfnamefont {A.}~\bibnamefont {Opala}}, \bibinfo {author} {\bibfnamefont {S.}~\bibnamefont {Ghosh}}, \bibinfo {author} {\bibfnamefont {L.}~\bibnamefont {Dominici}}, \bibinfo {author} {\bibfnamefont {V.}~\bibnamefont {Ardizzone}}, \bibinfo {author} {\bibfnamefont {M.}~\bibnamefont {De~Giorgi}}, \bibinfo {author} {\bibfnamefont {G.}~\bibnamefont {Lerario}}, \bibinfo {author} {\bibfnamefont {G.}~\bibnamefont {Gigli}}, \emph {et~al.},\ }\href@noop {} {\bibfield  {journal} {\bibinfo  {journal} {Nano Letters}\ }\textbf {\bibinfo {volume} {20}},\ \bibinfo {pages} {3506} (\bibinfo {year} {2020})}\BibitemShut {NoStop}%
\bibitem [{\citenamefont {Sedov}\ and\ \citenamefont {Kavokin}(2025)}]{sedov2025polariton}%
  \BibitemOpen
  \bibfield  {author} {\bibinfo {author} {\bibfnamefont {E.}~\bibnamefont {Sedov}}\ and\ \bibinfo {author} {\bibfnamefont {A.}~\bibnamefont {Kavokin}},\ }\href@noop {} {\bibfield  {journal} {\bibinfo  {journal} {Light: Science \& Applications}\ }\textbf {\bibinfo {volume} {14}},\ \bibinfo {pages} {52} (\bibinfo {year} {2025})}\BibitemShut {NoStop}%
\bibitem [{\citenamefont {Vahala}(2003)}]{Vahala2003a}%
  \BibitemOpen
  \bibfield  {author} {\bibinfo {author} {\bibfnamefont {K.~J.}\ \bibnamefont {Vahala}},\ }\href {https://doi.org/10.1038/nature01939} {\bibfield  {journal} {\bibinfo  {journal} {Nature}\ }\textbf {\bibinfo {volume} {424}},\ \bibinfo {pages} {839} (\bibinfo {year} {2003})}\BibitemShut {NoStop}%
\bibitem [{\citenamefont {Ma}\ \emph {et~al.}(2018)\citenamefont {Ma}, \citenamefont {{Alonso-Gonz{\'a}lez}}, \citenamefont {Li}, \citenamefont {Nikitin}, \citenamefont {Yuan}, \citenamefont {{Mart{\'i}n-S{\'a}nchez}}, \citenamefont {{Taboada-Guti{\'e}rrez}}, \citenamefont {Amenabar}, \citenamefont {Li}, \citenamefont {V{\'e}lez}, \citenamefont {Tollan}, \citenamefont {Dai}, \citenamefont {Zhang}, \citenamefont {Sriram}, \citenamefont {{Kalantar-Zadeh}}, \citenamefont {Lee}, \citenamefont {Hillenbrand},\ and\ \citenamefont {Bao}}]{Ma2018b}%
  \BibitemOpen
  \bibfield  {author} {\bibinfo {author} {\bibfnamefont {W.}~\bibnamefont {Ma}}, \bibinfo {author} {\bibfnamefont {P.}~\bibnamefont {{Alonso-Gonz{\'a}lez}}}, \bibinfo {author} {\bibfnamefont {S.}~\bibnamefont {Li}}, \bibinfo {author} {\bibfnamefont {A.~Y.}\ \bibnamefont {Nikitin}}, \bibinfo {author} {\bibfnamefont {J.}~\bibnamefont {Yuan}}, \bibinfo {author} {\bibfnamefont {J.}~\bibnamefont {{Mart{\'i}n-S{\'a}nchez}}}, \bibinfo {author} {\bibfnamefont {J.}~\bibnamefont {{Taboada-Guti{\'e}rrez}}}, \bibinfo {author} {\bibfnamefont {I.}~\bibnamefont {Amenabar}}, \bibinfo {author} {\bibfnamefont {P.}~\bibnamefont {Li}}, \bibinfo {author} {\bibfnamefont {S.}~\bibnamefont {V{\'e}lez}}, \bibinfo {author} {\bibfnamefont {C.}~\bibnamefont {Tollan}}, \bibinfo {author} {\bibfnamefont {Z.}~\bibnamefont {Dai}}, \bibinfo {author} {\bibfnamefont {Y.}~\bibnamefont {Zhang}}, \bibinfo {author} {\bibfnamefont {S.}~\bibnamefont {Sriram}}, \bibinfo {author} {\bibfnamefont {K.}~\bibnamefont {{Kalantar-Zadeh}}}, \bibinfo {author}
  {\bibfnamefont {S.-T.}\ \bibnamefont {Lee}}, \bibinfo {author} {\bibfnamefont {R.}~\bibnamefont {Hillenbrand}},\ and\ \bibinfo {author} {\bibfnamefont {Q.}~\bibnamefont {Bao}},\ }\href {https://doi.org/10.1038/s41586-018-0618-9} {\bibfield  {journal} {\bibinfo  {journal} {Nature}\ }\textbf {\bibinfo {volume} {562}},\ \bibinfo {pages} {557} (\bibinfo {year} {2018})}\BibitemShut {NoStop}%
\bibitem [{\citenamefont {{Taboada-Guti{\'e}rrez}}\ \emph {et~al.}(2020)\citenamefont {{Taboada-Guti{\'e}rrez}}, \citenamefont {{\'A}lvarez-P{\'e}rez}, \citenamefont {Duan}, \citenamefont {Ma}, \citenamefont {Crowley}, \citenamefont {Prieto}, \citenamefont {Bylinkin}, \citenamefont {Autore}, \citenamefont {Volkova}, \citenamefont {Kimura}, \citenamefont {Kimura}, \citenamefont {Berger}, \citenamefont {Li}, \citenamefont {Bao}, \citenamefont {Gao}, \citenamefont {Errea}, \citenamefont {Nikitin}, \citenamefont {Hillenbrand}, \citenamefont {{Mart{\'i}n-S{\'a}nchez}},\ and\ \citenamefont {{Alonso-Gonz{\'a}lez}}}]{Taboada-Gutierrez2020}%
  \BibitemOpen
  \bibfield  {author} {\bibinfo {author} {\bibfnamefont {J.}~\bibnamefont {{Taboada-Guti{\'e}rrez}}}, \bibinfo {author} {\bibfnamefont {G.}~\bibnamefont {{\'A}lvarez-P{\'e}rez}}, \bibinfo {author} {\bibfnamefont {J.}~\bibnamefont {Duan}}, \bibinfo {author} {\bibfnamefont {W.}~\bibnamefont {Ma}}, \bibinfo {author} {\bibfnamefont {K.}~\bibnamefont {Crowley}}, \bibinfo {author} {\bibfnamefont {I.}~\bibnamefont {Prieto}}, \bibinfo {author} {\bibfnamefont {A.}~\bibnamefont {Bylinkin}}, \bibinfo {author} {\bibfnamefont {M.}~\bibnamefont {Autore}}, \bibinfo {author} {\bibfnamefont {H.}~\bibnamefont {Volkova}}, \bibinfo {author} {\bibfnamefont {K.}~\bibnamefont {Kimura}}, \bibinfo {author} {\bibfnamefont {T.}~\bibnamefont {Kimura}}, \bibinfo {author} {\bibfnamefont {M.-H.}\ \bibnamefont {Berger}}, \bibinfo {author} {\bibfnamefont {S.}~\bibnamefont {Li}}, \bibinfo {author} {\bibfnamefont {Q.}~\bibnamefont {Bao}}, \bibinfo {author} {\bibfnamefont {X.~P.~A.}\ \bibnamefont {Gao}}, \bibinfo {author} {\bibfnamefont
  {I.}~\bibnamefont {Errea}}, \bibinfo {author} {\bibfnamefont {A.~Y.}\ \bibnamefont {Nikitin}}, \bibinfo {author} {\bibfnamefont {R.}~\bibnamefont {Hillenbrand}}, \bibinfo {author} {\bibfnamefont {J.}~\bibnamefont {{Mart{\'i}n-S{\'a}nchez}}},\ and\ \bibinfo {author} {\bibfnamefont {P.}~\bibnamefont {{Alonso-Gonz{\'a}lez}}},\ }\href {https://doi.org/10.1038/s41563-020-0665-0} {\bibfield  {journal} {\bibinfo  {journal} {Nature Materials}\ }\textbf {\bibinfo {volume} {19}},\ \bibinfo {pages} {964} (\bibinfo {year} {2020})}\BibitemShut {NoStop}%
\bibitem [{\citenamefont {Ruta}\ \emph {et~al.}(2023)\citenamefont {Ruta}, \citenamefont {Zhang}, \citenamefont {Shao}, \citenamefont {Moore}, \citenamefont {Acharya}, \citenamefont {Sun}, \citenamefont {Qiu}, \citenamefont {Geurs}, \citenamefont {Kim}, \citenamefont {Fu}, \citenamefont {Chica}, \citenamefont {Pashov}, \citenamefont {Xu}, \citenamefont {Xiao}, \citenamefont {Delor}, \citenamefont {Zhu}, \citenamefont {Millis}, \citenamefont {Roy}, \citenamefont {Hone}, \citenamefont {Dean}, \citenamefont {Katsnelson}, \citenamefont {van Schilfgaarde},\ and\ \citenamefont {Basov}}]{ruta_hyperbolic_2023}%
  \BibitemOpen
  \bibfield  {author} {\bibinfo {author} {\bibfnamefont {F.~L.}\ \bibnamefont {Ruta}}, \bibinfo {author} {\bibfnamefont {S.}~\bibnamefont {Zhang}}, \bibinfo {author} {\bibfnamefont {Y.}~\bibnamefont {Shao}}, \bibinfo {author} {\bibfnamefont {S.~L.}\ \bibnamefont {Moore}}, \bibinfo {author} {\bibfnamefont {S.}~\bibnamefont {Acharya}}, \bibinfo {author} {\bibfnamefont {Z.}~\bibnamefont {Sun}}, \bibinfo {author} {\bibfnamefont {S.}~\bibnamefont {Qiu}}, \bibinfo {author} {\bibfnamefont {J.}~\bibnamefont {Geurs}}, \bibinfo {author} {\bibfnamefont {B.~S.~Y.}\ \bibnamefont {Kim}}, \bibinfo {author} {\bibfnamefont {M.}~\bibnamefont {Fu}}, \bibinfo {author} {\bibfnamefont {D.~G.}\ \bibnamefont {Chica}}, \bibinfo {author} {\bibfnamefont {D.}~\bibnamefont {Pashov}}, \bibinfo {author} {\bibfnamefont {X.}~\bibnamefont {Xu}}, \bibinfo {author} {\bibfnamefont {D.}~\bibnamefont {Xiao}}, \bibinfo {author} {\bibfnamefont {M.}~\bibnamefont {Delor}}, \bibinfo {author} {\bibfnamefont {X.-Y.}\ \bibnamefont {Zhu}}, \bibinfo
  {author} {\bibfnamefont {A.~J.}\ \bibnamefont {Millis}}, \bibinfo {author} {\bibfnamefont {X.}~\bibnamefont {Roy}}, \bibinfo {author} {\bibfnamefont {J.~C.}\ \bibnamefont {Hone}}, \bibinfo {author} {\bibfnamefont {C.~R.}\ \bibnamefont {Dean}}, \bibinfo {author} {\bibfnamefont {M.~I.}\ \bibnamefont {Katsnelson}}, \bibinfo {author} {\bibfnamefont {M.}~\bibnamefont {van Schilfgaarde}},\ and\ \bibinfo {author} {\bibfnamefont {D.~N.}\ \bibnamefont {Basov}},\ }\href {https://doi.org/10.1038/s41467-023-44100-6} {\bibfield  {journal} {\bibinfo  {journal} {Nature Communications}\ }\textbf {\bibinfo {volume} {14}},\ \bibinfo {pages} {8261} (\bibinfo {year} {2023})}\BibitemShut {NoStop}%
\bibitem [{\citenamefont {Telford}\ \emph {et~al.}(2020)\citenamefont {Telford}, \citenamefont {Dismukes}, \citenamefont {Lee}, \citenamefont {Cheng}, \citenamefont {Wieteska}, \citenamefont {Bartholomew}, \citenamefont {Chen}, \citenamefont {Xu}, \citenamefont {Pasupathy}, \citenamefont {Zhu}, \citenamefont {Dean},\ and\ \citenamefont {Roy}}]{telford_layered_2020}%
  \BibitemOpen
  \bibfield  {author} {\bibinfo {author} {\bibfnamefont {E.~J.}\ \bibnamefont {Telford}}, \bibinfo {author} {\bibfnamefont {A.~H.}\ \bibnamefont {Dismukes}}, \bibinfo {author} {\bibfnamefont {K.}~\bibnamefont {Lee}}, \bibinfo {author} {\bibfnamefont {M.}~\bibnamefont {Cheng}}, \bibinfo {author} {\bibfnamefont {A.}~\bibnamefont {Wieteska}}, \bibinfo {author} {\bibfnamefont {A.~K.}\ \bibnamefont {Bartholomew}}, \bibinfo {author} {\bibfnamefont {Y.-S.}\ \bibnamefont {Chen}}, \bibinfo {author} {\bibfnamefont {X.}~\bibnamefont {Xu}}, \bibinfo {author} {\bibfnamefont {A.~N.}\ \bibnamefont {Pasupathy}}, \bibinfo {author} {\bibfnamefont {X.}~\bibnamefont {Zhu}}, \bibinfo {author} {\bibfnamefont {C.~R.}\ \bibnamefont {Dean}},\ and\ \bibinfo {author} {\bibfnamefont {X.}~\bibnamefont {Roy}},\ }\href {https://doi.org/10.1002/adma.202003240} {\bibfield  {journal} {\bibinfo  {journal} {Advanced Materials}\ }\textbf {\bibinfo {volume} {32}},\ \bibinfo {pages} {2003240} (\bibinfo {year} {2020})}\BibitemShut {NoStop}%
\bibitem [{\citenamefont {Wang}\ \emph {et~al.}(2020)\citenamefont {Wang}, \citenamefont {Qi},\ and\ \citenamefont {Qian}}]{wang_electrically-tunable_2020}%
  \BibitemOpen
  \bibfield  {author} {\bibinfo {author} {\bibfnamefont {H.}~\bibnamefont {Wang}}, \bibinfo {author} {\bibfnamefont {J.}~\bibnamefont {Qi}},\ and\ \bibinfo {author} {\bibfnamefont {X.}~\bibnamefont {Qian}},\ }\href {https://doi.org/10.1063/5.0014865} {\bibfield  {journal} {\bibinfo  {journal} {Applied Physics Letters}\ }\textbf {\bibinfo {volume} {117}},\ \bibinfo {pages} {083102} (\bibinfo {year} {2020})}\BibitemShut {NoStop}%
\bibitem [{\citenamefont {Lee}\ \emph {et~al.}(2021)\citenamefont {Lee}, \citenamefont {Dismukes}, \citenamefont {Telford}, \citenamefont {Wiscons}, \citenamefont {Wang}, \citenamefont {Xu}, \citenamefont {Nuckolls}, \citenamefont {Dean}, \citenamefont {Roy},\ and\ \citenamefont {Zhu}}]{lee_magnetic_2021}%
  \BibitemOpen
  \bibfield  {author} {\bibinfo {author} {\bibfnamefont {K.}~\bibnamefont {Lee}}, \bibinfo {author} {\bibfnamefont {A.~H.}\ \bibnamefont {Dismukes}}, \bibinfo {author} {\bibfnamefont {E.~J.}\ \bibnamefont {Telford}}, \bibinfo {author} {\bibfnamefont {R.~A.}\ \bibnamefont {Wiscons}}, \bibinfo {author} {\bibfnamefont {J.}~\bibnamefont {Wang}}, \bibinfo {author} {\bibfnamefont {X.}~\bibnamefont {Xu}}, \bibinfo {author} {\bibfnamefont {C.}~\bibnamefont {Nuckolls}}, \bibinfo {author} {\bibfnamefont {C.~R.}\ \bibnamefont {Dean}}, \bibinfo {author} {\bibfnamefont {X.}~\bibnamefont {Roy}},\ and\ \bibinfo {author} {\bibfnamefont {X.}~\bibnamefont {Zhu}},\ }\href {https://doi.org/10.1021/acs.nanolett.1c00219} {\bibfield  {journal} {\bibinfo  {journal} {Nano Letters}\ }\textbf {\bibinfo {volume} {21}},\ \bibinfo {pages} {3511} (\bibinfo {year} {2021})}\BibitemShut {NoStop}%
\bibitem [{\citenamefont {Scheie}\ \emph {et~al.}(2022)\citenamefont {Scheie}, \citenamefont {Ziebel}, \citenamefont {Chica}, \citenamefont {Bae}, \citenamefont {Wang}, \citenamefont {Kolesnikov}, \citenamefont {Zhu},\ and\ \citenamefont {Roy}}]{scheie_spin_2022}%
  \BibitemOpen
  \bibfield  {author} {\bibinfo {author} {\bibfnamefont {A.}~\bibnamefont {Scheie}}, \bibinfo {author} {\bibfnamefont {M.}~\bibnamefont {Ziebel}}, \bibinfo {author} {\bibfnamefont {D.~G.}\ \bibnamefont {Chica}}, \bibinfo {author} {\bibfnamefont {Y.~J.}\ \bibnamefont {Bae}}, \bibinfo {author} {\bibfnamefont {X.}~\bibnamefont {Wang}}, \bibinfo {author} {\bibfnamefont {A.~I.}\ \bibnamefont {Kolesnikov}}, \bibinfo {author} {\bibfnamefont {X.}~\bibnamefont {Zhu}},\ and\ \bibinfo {author} {\bibfnamefont {X.}~\bibnamefont {Roy}},\ }\href {https://doi.org/10.1002/advs.202202467} {\bibfield  {journal} {\bibinfo  {journal} {Advanced Science}\ }\textbf {\bibinfo {volume} {9}},\ \bibinfo {pages} {2202467} (\bibinfo {year} {2022})}\BibitemShut {NoStop}%
\bibitem [{\citenamefont {Wu}\ \emph {et~al.}(2022)\citenamefont {Wu}, \citenamefont {Gutiérrez-Lezama}, \citenamefont {López-Paz}, \citenamefont {Gibertini}, \citenamefont {Watanabe}, \citenamefont {Taniguchi}, \citenamefont {von Rohr}, \citenamefont {Ubrig},\ and\ \citenamefont {Morpurgo}}]{wu_quasi-1d_2022}%
  \BibitemOpen
  \bibfield  {author} {\bibinfo {author} {\bibfnamefont {F.}~\bibnamefont {Wu}}, \bibinfo {author} {\bibfnamefont {I.}~\bibnamefont {Gutiérrez-Lezama}}, \bibinfo {author} {\bibfnamefont {S.~A.}\ \bibnamefont {López-Paz}}, \bibinfo {author} {\bibfnamefont {M.}~\bibnamefont {Gibertini}}, \bibinfo {author} {\bibfnamefont {K.}~\bibnamefont {Watanabe}}, \bibinfo {author} {\bibfnamefont {T.}~\bibnamefont {Taniguchi}}, \bibinfo {author} {\bibfnamefont {F.~O.}\ \bibnamefont {von Rohr}}, \bibinfo {author} {\bibfnamefont {N.}~\bibnamefont {Ubrig}},\ and\ \bibinfo {author} {\bibfnamefont {A.~F.}\ \bibnamefont {Morpurgo}},\ }\href {https://doi.org/10.1002/adma.202109759} {\bibfield  {journal} {\bibinfo  {journal} {Advanced Materials}\ }\textbf {\bibinfo {volume} {34}},\ \bibinfo {pages} {2109759} (\bibinfo {year} {2022})}\BibitemShut {NoStop}%
\bibitem [{\citenamefont {Klein}\ \emph {et~al.}(2023)\citenamefont {Klein}, \citenamefont {Song}, \citenamefont {Pingault}, \citenamefont {Dirnberger}, \citenamefont {Chi}, \citenamefont {Curtis}, \citenamefont {Dana}, \citenamefont {Bushati}, \citenamefont {Quan}, \citenamefont {Dekanovsky}, \citenamefont {Sofer}, \citenamefont {Alù}, \citenamefont {Menon}, \citenamefont {Moodera}, \citenamefont {Lončar}, \citenamefont {Narang},\ and\ \citenamefont {Ross}}]{klein_sensing_2023}%
  \BibitemOpen
  \bibfield  {author} {\bibinfo {author} {\bibfnamefont {J.}~\bibnamefont {Klein}}, \bibinfo {author} {\bibfnamefont {Z.}~\bibnamefont {Song}}, \bibinfo {author} {\bibfnamefont {B.}~\bibnamefont {Pingault}}, \bibinfo {author} {\bibfnamefont {F.}~\bibnamefont {Dirnberger}}, \bibinfo {author} {\bibfnamefont {H.}~\bibnamefont {Chi}}, \bibinfo {author} {\bibfnamefont {J.~B.}\ \bibnamefont {Curtis}}, \bibinfo {author} {\bibfnamefont {R.}~\bibnamefont {Dana}}, \bibinfo {author} {\bibfnamefont {R.}~\bibnamefont {Bushati}}, \bibinfo {author} {\bibfnamefont {J.}~\bibnamefont {Quan}}, \bibinfo {author} {\bibfnamefont {L.}~\bibnamefont {Dekanovsky}}, \bibinfo {author} {\bibfnamefont {Z.}~\bibnamefont {Sofer}}, \bibinfo {author} {\bibfnamefont {A.}~\bibnamefont {Alù}}, \bibinfo {author} {\bibfnamefont {V.~M.}\ \bibnamefont {Menon}}, \bibinfo {author} {\bibfnamefont {J.~S.}\ \bibnamefont {Moodera}}, \bibinfo {author} {\bibfnamefont {M.}~\bibnamefont {Lončar}}, \bibinfo {author} {\bibfnamefont {P.}~\bibnamefont
  {Narang}},\ and\ \bibinfo {author} {\bibfnamefont {F.~M.}\ \bibnamefont {Ross}},\ }\href {https://doi.org/10.1021/acsnano.2c07655} {\bibfield  {journal} {\bibinfo  {journal} {ACS Nano}\ }\textbf {\bibinfo {volume} {17}},\ \bibinfo {pages} {288} (\bibinfo {year} {2023})}\BibitemShut {NoStop}%
\bibitem [{\citenamefont {Wilson}\ \emph {et~al.}(2021)\citenamefont {Wilson}, \citenamefont {Lee}, \citenamefont {Cenker}, \citenamefont {Xie}, \citenamefont {Dismukes}, \citenamefont {Telford}, \citenamefont {Fonseca}, \citenamefont {Sivakumar}, \citenamefont {Dean}, \citenamefont {Cao}, \citenamefont {Roy}, \citenamefont {Xu},\ and\ \citenamefont {Zhu}}]{wilson_interlayer_2021}%
  \BibitemOpen
  \bibfield  {author} {\bibinfo {author} {\bibfnamefont {N.~P.}\ \bibnamefont {Wilson}}, \bibinfo {author} {\bibfnamefont {K.}~\bibnamefont {Lee}}, \bibinfo {author} {\bibfnamefont {J.}~\bibnamefont {Cenker}}, \bibinfo {author} {\bibfnamefont {K.}~\bibnamefont {Xie}}, \bibinfo {author} {\bibfnamefont {A.~H.}\ \bibnamefont {Dismukes}}, \bibinfo {author} {\bibfnamefont {E.~J.}\ \bibnamefont {Telford}}, \bibinfo {author} {\bibfnamefont {J.}~\bibnamefont {Fonseca}}, \bibinfo {author} {\bibfnamefont {S.}~\bibnamefont {Sivakumar}}, \bibinfo {author} {\bibfnamefont {C.}~\bibnamefont {Dean}}, \bibinfo {author} {\bibfnamefont {T.}~\bibnamefont {Cao}}, \bibinfo {author} {\bibfnamefont {X.}~\bibnamefont {Roy}}, \bibinfo {author} {\bibfnamefont {X.}~\bibnamefont {Xu}},\ and\ \bibinfo {author} {\bibfnamefont {X.}~\bibnamefont {Zhu}},\ }\href {https://doi.org/10.1038/s41563-021-01070-8} {\bibfield  {journal} {\bibinfo  {journal} {Nature Materials}\ }\textbf {\bibinfo {volume} {20}},\ \bibinfo {pages} {1657} (\bibinfo
  {year} {2021})}\BibitemShut {NoStop}%
\bibitem [{\citenamefont {Bae}\ \emph {et~al.}(2022)\citenamefont {Bae}, \citenamefont {Wang}, \citenamefont {Scheie}, \citenamefont {Xu}, \citenamefont {Chica}, \citenamefont {Diederich}, \citenamefont {Cenker}, \citenamefont {Ziebel}, \citenamefont {Bai}, \citenamefont {Ren}, \citenamefont {Dean}, \citenamefont {Delor}, \citenamefont {Xu}, \citenamefont {Roy}, \citenamefont {Kent},\ and\ \citenamefont {Zhu}}]{bae_exciton-coupled_2022}%
  \BibitemOpen
  \bibfield  {author} {\bibinfo {author} {\bibfnamefont {Y.~J.}\ \bibnamefont {Bae}}, \bibinfo {author} {\bibfnamefont {J.}~\bibnamefont {Wang}}, \bibinfo {author} {\bibfnamefont {A.}~\bibnamefont {Scheie}}, \bibinfo {author} {\bibfnamefont {J.}~\bibnamefont {Xu}}, \bibinfo {author} {\bibfnamefont {D.~G.}\ \bibnamefont {Chica}}, \bibinfo {author} {\bibfnamefont {G.~M.}\ \bibnamefont {Diederich}}, \bibinfo {author} {\bibfnamefont {J.}~\bibnamefont {Cenker}}, \bibinfo {author} {\bibfnamefont {M.~E.}\ \bibnamefont {Ziebel}}, \bibinfo {author} {\bibfnamefont {Y.}~\bibnamefont {Bai}}, \bibinfo {author} {\bibfnamefont {H.}~\bibnamefont {Ren}}, \bibinfo {author} {\bibfnamefont {C.~R.}\ \bibnamefont {Dean}}, \bibinfo {author} {\bibfnamefont {M.}~\bibnamefont {Delor}}, \bibinfo {author} {\bibfnamefont {X.}~\bibnamefont {Xu}}, \bibinfo {author} {\bibfnamefont {X.}~\bibnamefont {Roy}}, \bibinfo {author} {\bibfnamefont {A.~D.}\ \bibnamefont {Kent}},\ and\ \bibinfo {author} {\bibfnamefont {X.}~\bibnamefont {Zhu}},\ }\href
  {https://doi.org/10.1038/s41586-022-05024-1} {\bibfield  {journal} {\bibinfo  {journal} {Nature}\ }\textbf {\bibinfo {volume} {609}},\ \bibinfo {pages} {282} (\bibinfo {year} {2022})}\BibitemShut {NoStop}%
\bibitem [{\citenamefont {Diederich}\ \emph {et~al.}(2023)\citenamefont {Diederich}, \citenamefont {Cenker}, \citenamefont {Ren}, \citenamefont {Fonseca}, \citenamefont {Chica}, \citenamefont {Bae}, \citenamefont {Zhu}, \citenamefont {Roy}, \citenamefont {Cao}, \citenamefont {Xiao},\ and\ \citenamefont {Xu}}]{diederich_tunable_2023}%
  \BibitemOpen
  \bibfield  {author} {\bibinfo {author} {\bibfnamefont {G.~M.}\ \bibnamefont {Diederich}}, \bibinfo {author} {\bibfnamefont {J.}~\bibnamefont {Cenker}}, \bibinfo {author} {\bibfnamefont {Y.}~\bibnamefont {Ren}}, \bibinfo {author} {\bibfnamefont {J.}~\bibnamefont {Fonseca}}, \bibinfo {author} {\bibfnamefont {D.~G.}\ \bibnamefont {Chica}}, \bibinfo {author} {\bibfnamefont {Y.~J.}\ \bibnamefont {Bae}}, \bibinfo {author} {\bibfnamefont {X.}~\bibnamefont {Zhu}}, \bibinfo {author} {\bibfnamefont {X.}~\bibnamefont {Roy}}, \bibinfo {author} {\bibfnamefont {T.}~\bibnamefont {Cao}}, \bibinfo {author} {\bibfnamefont {D.}~\bibnamefont {Xiao}},\ and\ \bibinfo {author} {\bibfnamefont {X.}~\bibnamefont {Xu}},\ }\href {https://doi.org/10.1038/s41565-022-01259-1} {\bibfield  {journal} {\bibinfo  {journal} {Nature Nanotechnology}\ }\textbf {\bibinfo {volume} {18}},\ \bibinfo {pages} {23} (\bibinfo {year} {2023})}\BibitemShut {NoStop}%
\bibitem [{\citenamefont {Dirnberger}\ \emph {et~al.}(2023)\citenamefont {Dirnberger}, \citenamefont {Quan}, \citenamefont {Bushati}, \citenamefont {Diederich}, \citenamefont {Florian}, \citenamefont {Klein}, \citenamefont {Mosina}, \citenamefont {Sofer}, \citenamefont {Xu}, \citenamefont {Kamra}, \citenamefont {García-Vidal}, \citenamefont {Alù},\ and\ \citenamefont {Menon}}]{dirnberger_magneto-optics_2023}%
  \BibitemOpen
  \bibfield  {author} {\bibinfo {author} {\bibfnamefont {F.}~\bibnamefont {Dirnberger}}, \bibinfo {author} {\bibfnamefont {J.}~\bibnamefont {Quan}}, \bibinfo {author} {\bibfnamefont {R.}~\bibnamefont {Bushati}}, \bibinfo {author} {\bibfnamefont {G.~M.}\ \bibnamefont {Diederich}}, \bibinfo {author} {\bibfnamefont {M.}~\bibnamefont {Florian}}, \bibinfo {author} {\bibfnamefont {J.}~\bibnamefont {Klein}}, \bibinfo {author} {\bibfnamefont {K.}~\bibnamefont {Mosina}}, \bibinfo {author} {\bibfnamefont {Z.}~\bibnamefont {Sofer}}, \bibinfo {author} {\bibfnamefont {X.}~\bibnamefont {Xu}}, \bibinfo {author} {\bibfnamefont {A.}~\bibnamefont {Kamra}}, \bibinfo {author} {\bibfnamefont {F.~J.}\ \bibnamefont {García-Vidal}}, \bibinfo {author} {\bibfnamefont {A.}~\bibnamefont {Alù}},\ and\ \bibinfo {author} {\bibfnamefont {V.~M.}\ \bibnamefont {Menon}},\ }\href {https://doi.org/10.1038/s41586-023-06275-2} {\bibfield  {journal} {\bibinfo  {journal} {Nature}\ }\textbf {\bibinfo {volume} {620}},\ \bibinfo {pages} {533}
  (\bibinfo {year} {2023})}\BibitemShut {NoStop}%
\bibitem [{\citenamefont {Ziebel}\ \emph {et~al.}(2024)\citenamefont {Ziebel}, \citenamefont {Feuer}, \citenamefont {Cox}, \citenamefont {Zhu}, \citenamefont {Dean},\ and\ \citenamefont {Roy}}]{ziebel2024crsbr}%
  \BibitemOpen
  \bibfield  {author} {\bibinfo {author} {\bibfnamefont {M.~E.}\ \bibnamefont {Ziebel}}, \bibinfo {author} {\bibfnamefont {M.~L.}\ \bibnamefont {Feuer}}, \bibinfo {author} {\bibfnamefont {J.}~\bibnamefont {Cox}}, \bibinfo {author} {\bibfnamefont {X.}~\bibnamefont {Zhu}}, \bibinfo {author} {\bibfnamefont {C.~R.}\ \bibnamefont {Dean}},\ and\ \bibinfo {author} {\bibfnamefont {X.}~\bibnamefont {Roy}},\ }\href@noop {} {\bibfield  {journal} {\bibinfo  {journal} {Nano Letters}\ }\textbf {\bibinfo {volume} {24}},\ \bibinfo {pages} {4319} (\bibinfo {year} {2024})}\BibitemShut {NoStop}%
\bibitem [{\citenamefont {Datta}\ \emph {et~al.}(2025)\citenamefont {Datta}, \citenamefont {Adak}, \citenamefont {Yu}, \citenamefont {Valiyaparambil~Dharmapalan}, \citenamefont {Hall}, \citenamefont {Vakulenko}, \citenamefont {Komissarenko}, \citenamefont {Kurganov}, \citenamefont {Quan}, \citenamefont {Wang} \emph {et~al.}}]{datta2025magnon}%
  \BibitemOpen
  \bibfield  {author} {\bibinfo {author} {\bibfnamefont {B.}~\bibnamefont {Datta}}, \bibinfo {author} {\bibfnamefont {P.~C.}\ \bibnamefont {Adak}}, \bibinfo {author} {\bibfnamefont {S.}~\bibnamefont {Yu}}, \bibinfo {author} {\bibfnamefont {A.}~\bibnamefont {Valiyaparambil~Dharmapalan}}, \bibinfo {author} {\bibfnamefont {S.~J.}\ \bibnamefont {Hall}}, \bibinfo {author} {\bibfnamefont {A.}~\bibnamefont {Vakulenko}}, \bibinfo {author} {\bibfnamefont {F.}~\bibnamefont {Komissarenko}}, \bibinfo {author} {\bibfnamefont {E.}~\bibnamefont {Kurganov}}, \bibinfo {author} {\bibfnamefont {J.}~\bibnamefont {Quan}}, \bibinfo {author} {\bibfnamefont {W.}~\bibnamefont {Wang}}, \emph {et~al.},\ }\href@noop {} {\bibfield  {journal} {\bibinfo  {journal} {Nature Materials}\ ,\ \bibinfo {pages} {1}} (\bibinfo {year} {2025})}\BibitemShut {NoStop}%
\bibitem [{\citenamefont {Wang}\ \emph {et~al.}(2023)\citenamefont {Wang}, \citenamefont {Zhang}, \citenamefont {Yang}, \citenamefont {Lin}, \citenamefont {Chen}, \citenamefont {Yang}, \citenamefont {Gong}, \citenamefont {Chen}, \citenamefont {Ye},\ and\ \citenamefont {Liu}}]{wang_magnetically-dressed_2023}%
  \BibitemOpen
  \bibfield  {author} {\bibinfo {author} {\bibfnamefont {T.}~\bibnamefont {Wang}}, \bibinfo {author} {\bibfnamefont {D.}~\bibnamefont {Zhang}}, \bibinfo {author} {\bibfnamefont {S.}~\bibnamefont {Yang}}, \bibinfo {author} {\bibfnamefont {Z.}~\bibnamefont {Lin}}, \bibinfo {author} {\bibfnamefont {Q.}~\bibnamefont {Chen}}, \bibinfo {author} {\bibfnamefont {J.}~\bibnamefont {Yang}}, \bibinfo {author} {\bibfnamefont {Q.}~\bibnamefont {Gong}}, \bibinfo {author} {\bibfnamefont {Z.}~\bibnamefont {Chen}}, \bibinfo {author} {\bibfnamefont {Y.}~\bibnamefont {Ye}},\ and\ \bibinfo {author} {\bibfnamefont {W.}~\bibnamefont {Liu}},\ }\href {https://doi.org/10.1038/s41467-023-41688-7} {\bibfield  {journal} {\bibinfo  {journal} {Nature Communications}\ }\textbf {\bibinfo {volume} {14}},\ \bibinfo {pages} {5966} (\bibinfo {year} {2023})}\BibitemShut {NoStop}%
\bibitem [{\citenamefont {Li}\ \emph {et~al.}(2024{\natexlab{a}})\citenamefont {Li}, \citenamefont {Shen}, \citenamefont {Jiang}, \citenamefont {Tang}, \citenamefont {Liu}, \citenamefont {Guo}, \citenamefont {Liang}, \citenamefont {Song}, \citenamefont {Deng},\ and\ \citenamefont {Zhang}}]{Li2024e}%
  \BibitemOpen
  \bibfield  {author} {\bibinfo {author} {\bibfnamefont {C.}~\bibnamefont {Li}}, \bibinfo {author} {\bibfnamefont {C.}~\bibnamefont {Shen}}, \bibinfo {author} {\bibfnamefont {N.}~\bibnamefont {Jiang}}, \bibinfo {author} {\bibfnamefont {K.~K.}\ \bibnamefont {Tang}}, \bibinfo {author} {\bibfnamefont {X.}~\bibnamefont {Liu}}, \bibinfo {author} {\bibfnamefont {J.}~\bibnamefont {Guo}}, \bibinfo {author} {\bibfnamefont {Y.}~\bibnamefont {Liang}}, \bibinfo {author} {\bibfnamefont {J.}~\bibnamefont {Song}}, \bibinfo {author} {\bibfnamefont {X.}~\bibnamefont {Deng}},\ and\ \bibinfo {author} {\bibfnamefont {Q.}~\bibnamefont {Zhang}},\ }\href {https://doi.org/10.1002/adfm.202411589} {\bibfield  {journal} {\bibinfo  {journal} {Advanced Functional Materials}\ }\textbf {\bibinfo {volume} {34}},\ \bibinfo {pages} {2411589} (\bibinfo {year} {2024}{\natexlab{a}})}\BibitemShut {NoStop}%
\bibitem [{\citenamefont {Tabataba-Vakili}\ \emph {et~al.}(2024)\citenamefont {Tabataba-Vakili}, \citenamefont {Nguyen}, \citenamefont {Rupp}, \citenamefont {Mosina}, \citenamefont {Papavasileiou}, \citenamefont {Watanabe}, \citenamefont {Taniguchi}, \citenamefont {Maletinsky}, \citenamefont {Glazov}, \citenamefont {Sofer}, \citenamefont {Baimuratov},\ and\ \citenamefont {Högele}}]{tabataba-vakili_doping-control_2024}%
  \BibitemOpen
  \bibfield  {author} {\bibinfo {author} {\bibfnamefont {F.}~\bibnamefont {Tabataba-Vakili}}, \bibinfo {author} {\bibfnamefont {H.~P.~G.}\ \bibnamefont {Nguyen}}, \bibinfo {author} {\bibfnamefont {A.}~\bibnamefont {Rupp}}, \bibinfo {author} {\bibfnamefont {K.}~\bibnamefont {Mosina}}, \bibinfo {author} {\bibfnamefont {A.}~\bibnamefont {Papavasileiou}}, \bibinfo {author} {\bibfnamefont {K.}~\bibnamefont {Watanabe}}, \bibinfo {author} {\bibfnamefont {T.}~\bibnamefont {Taniguchi}}, \bibinfo {author} {\bibfnamefont {P.}~\bibnamefont {Maletinsky}}, \bibinfo {author} {\bibfnamefont {M.~M.}\ \bibnamefont {Glazov}}, \bibinfo {author} {\bibfnamefont {Z.}~\bibnamefont {Sofer}}, \bibinfo {author} {\bibfnamefont {A.~S.}\ \bibnamefont {Baimuratov}},\ and\ \bibinfo {author} {\bibfnamefont {A.}~\bibnamefont {Högele}},\ }\href {https://doi.org/10.1038/s41467-024-49048-9} {\bibfield  {journal} {\bibinfo  {journal} {Nature Communications}\ }\textbf {\bibinfo {volume} {15}},\ \bibinfo {pages} {4735} (\bibinfo {year}
  {2024})}\BibitemShut {NoStop}%
\bibitem [{\citenamefont {Brennan}\ \emph {et~al.}(2024)\citenamefont {Brennan}, \citenamefont {Noble}, \citenamefont {Tang}, \citenamefont {Ziebel},\ and\ \citenamefont {Bae}}]{brennan_important_2024}%
  \BibitemOpen
  \bibfield  {author} {\bibinfo {author} {\bibfnamefont {N.~J.}\ \bibnamefont {Brennan}}, \bibinfo {author} {\bibfnamefont {C.~A.}\ \bibnamefont {Noble}}, \bibinfo {author} {\bibfnamefont {J.}~\bibnamefont {Tang}}, \bibinfo {author} {\bibfnamefont {M.~E.}\ \bibnamefont {Ziebel}},\ and\ \bibinfo {author} {\bibfnamefont {Y.~J.}\ \bibnamefont {Bae}},\ }\href {https://doi.org/10.1021/acsphyschemau.4c00010} {\bibfield  {journal} {\bibinfo  {journal} {ACS Physical Chemistry Au}\ }\textbf {\bibinfo {volume} {4}},\ \bibinfo {pages} {322} (\bibinfo {year} {2024})}\BibitemShut {NoStop}%
\bibitem [{\citenamefont {Li}\ \emph {et~al.}(2024{\natexlab{b}})\citenamefont {Li}, \citenamefont {Xie}, \citenamefont {Alfrey}, \citenamefont {Beach}, \citenamefont {McLellan}, \citenamefont {Lu}, \citenamefont {Hu}, \citenamefont {Liu}, \citenamefont {Dhale}, \citenamefont {Lv}, \citenamefont {Zhao}, \citenamefont {Sun},\ and\ \citenamefont {Deng}}]{Li2024f}%
  \BibitemOpen
  \bibfield  {author} {\bibinfo {author} {\bibfnamefont {Q.}~\bibnamefont {Li}}, \bibinfo {author} {\bibfnamefont {X.}~\bibnamefont {Xie}}, \bibinfo {author} {\bibfnamefont {A.}~\bibnamefont {Alfrey}}, \bibinfo {author} {\bibfnamefont {C.~W.}\ \bibnamefont {Beach}}, \bibinfo {author} {\bibfnamefont {N.}~\bibnamefont {McLellan}}, \bibinfo {author} {\bibfnamefont {Y.}~\bibnamefont {Lu}}, \bibinfo {author} {\bibfnamefont {J.}~\bibnamefont {Hu}}, \bibinfo {author} {\bibfnamefont {W.}~\bibnamefont {Liu}}, \bibinfo {author} {\bibfnamefont {N.}~\bibnamefont {Dhale}}, \bibinfo {author} {\bibfnamefont {B.}~\bibnamefont {Lv}}, \bibinfo {author} {\bibfnamefont {L.}~\bibnamefont {Zhao}}, \bibinfo {author} {\bibfnamefont {K.}~\bibnamefont {Sun}},\ and\ \bibinfo {author} {\bibfnamefont {H.}~\bibnamefont {Deng}},\ }\href {https://doi.org/10.1103/PhysRevLett.133.266901} {\bibfield  {journal} {\bibinfo  {journal} {Physical Review Letters}\ }\textbf {\bibinfo {volume} {133}},\ \bibinfo {pages} {266901} (\bibinfo {year}
  {2024}{\natexlab{b}})}\BibitemShut {NoStop}%
\bibitem [{\citenamefont {Guo}\ \emph {et~al.}(2018)\citenamefont {Guo}, \citenamefont {Zhang}, \citenamefont {Yuan}, \citenamefont {Wang},\ and\ \citenamefont {Wang}}]{guo_chromium_2018}%
  \BibitemOpen
  \bibfield  {author} {\bibinfo {author} {\bibfnamefont {Y.}~\bibnamefont {Guo}}, \bibinfo {author} {\bibfnamefont {Y.}~\bibnamefont {Zhang}}, \bibinfo {author} {\bibfnamefont {S.}~\bibnamefont {Yuan}}, \bibinfo {author} {\bibfnamefont {B.}~\bibnamefont {Wang}},\ and\ \bibinfo {author} {\bibfnamefont {J.}~\bibnamefont {Wang}},\ }\href {https://doi.org/10.1039/C8NR06368K} {\bibfield  {journal} {\bibinfo  {journal} {Nanoscale}\ }\textbf {\bibinfo {volume} {10}},\ \bibinfo {pages} {18036} (\bibinfo {year} {2018})}\BibitemShut {NoStop}%
\bibitem [{\citenamefont {Watson}\ \emph {et~al.}(2024)\citenamefont {Watson}, \citenamefont {Acharya}, \citenamefont {Nunn}, \citenamefont {Nagireddy}, \citenamefont {Pashov}, \citenamefont {Rösner}, \citenamefont {van Schilfgaarde}, \citenamefont {Wilson},\ and\ \citenamefont {Cacho}}]{watson2024giant}%
  \BibitemOpen
  \bibfield  {author} {\bibinfo {author} {\bibfnamefont {M.~D.}\ \bibnamefont {Watson}}, \bibinfo {author} {\bibfnamefont {S.}~\bibnamefont {Acharya}}, \bibinfo {author} {\bibfnamefont {J.~E.}\ \bibnamefont {Nunn}}, \bibinfo {author} {\bibfnamefont {L.}~\bibnamefont {Nagireddy}}, \bibinfo {author} {\bibfnamefont {D.}~\bibnamefont {Pashov}}, \bibinfo {author} {\bibfnamefont {M.}~\bibnamefont {Rösner}}, \bibinfo {author} {\bibfnamefont {M.}~\bibnamefont {van Schilfgaarde}}, \bibinfo {author} {\bibfnamefont {N.~R.}\ \bibnamefont {Wilson}},\ and\ \bibinfo {author} {\bibfnamefont {C.}~\bibnamefont {Cacho}},\ }\href {https://doi.org/10.1038/s41699-024-00492-7} {\bibfield  {journal} {\bibinfo  {journal} {npj 2D Materials and Applications}\ }\textbf {\bibinfo {volume} {8}},\ \bibinfo {pages} {1} (\bibinfo {year} {2024})}\BibitemShut {NoStop}%
\bibitem [{\citenamefont {Shao}\ \emph {et~al.}(2025)\citenamefont {Shao}, \citenamefont {Dirnberger}, \citenamefont {Qiu}, \citenamefont {Acharya}, \citenamefont {Terres}, \citenamefont {Telford}, \citenamefont {Pashov}, \citenamefont {Kim}, \citenamefont {Ruta}, \citenamefont {Chica}, \citenamefont {Dismukes}, \citenamefont {Ziebel}, \citenamefont {Wang}, \citenamefont {Choe}, \citenamefont {Bae}, \citenamefont {Millis}, \citenamefont {Katsnelson}, \citenamefont {Mosina}, \citenamefont {Sofer}, \citenamefont {Huber}, \citenamefont {Zhu}, \citenamefont {Roy}, \citenamefont {{van Schilfgaarde}}, \citenamefont {Chernikov},\ and\ \citenamefont {Basov}}]{shaoMagneticallyConfinedSurface2025}%
  \BibitemOpen
  \bibfield  {author} {\bibinfo {author} {\bibfnamefont {Y.}~\bibnamefont {Shao}}, \bibinfo {author} {\bibfnamefont {F.}~\bibnamefont {Dirnberger}}, \bibinfo {author} {\bibfnamefont {S.}~\bibnamefont {Qiu}}, \bibinfo {author} {\bibfnamefont {S.}~\bibnamefont {Acharya}}, \bibinfo {author} {\bibfnamefont {S.}~\bibnamefont {Terres}}, \bibinfo {author} {\bibfnamefont {E.~J.}\ \bibnamefont {Telford}}, \bibinfo {author} {\bibfnamefont {D.}~\bibnamefont {Pashov}}, \bibinfo {author} {\bibfnamefont {B.~S.~Y.}\ \bibnamefont {Kim}}, \bibinfo {author} {\bibfnamefont {F.~L.}\ \bibnamefont {Ruta}}, \bibinfo {author} {\bibfnamefont {D.~G.}\ \bibnamefont {Chica}}, \bibinfo {author} {\bibfnamefont {A.~H.}\ \bibnamefont {Dismukes}}, \bibinfo {author} {\bibfnamefont {M.~E.}\ \bibnamefont {Ziebel}}, \bibinfo {author} {\bibfnamefont {Y.}~\bibnamefont {Wang}}, \bibinfo {author} {\bibfnamefont {J.}~\bibnamefont {Choe}}, \bibinfo {author} {\bibfnamefont {Y.~J.}\ \bibnamefont {Bae}}, \bibinfo {author} {\bibfnamefont {A.~J.}\
  \bibnamefont {Millis}}, \bibinfo {author} {\bibfnamefont {M.~I.}\ \bibnamefont {Katsnelson}}, \bibinfo {author} {\bibfnamefont {K.}~\bibnamefont {Mosina}}, \bibinfo {author} {\bibfnamefont {Z.}~\bibnamefont {Sofer}}, \bibinfo {author} {\bibfnamefont {R.}~\bibnamefont {Huber}}, \bibinfo {author} {\bibfnamefont {X.}~\bibnamefont {Zhu}}, \bibinfo {author} {\bibfnamefont {X.}~\bibnamefont {Roy}}, \bibinfo {author} {\bibfnamefont {M.}~\bibnamefont {{van Schilfgaarde}}}, \bibinfo {author} {\bibfnamefont {A.}~\bibnamefont {Chernikov}},\ and\ \bibinfo {author} {\bibfnamefont {D.~N.}\ \bibnamefont {Basov}},\ }\bibfield  {journal} {\bibinfo  {journal} {Nature Materials}\ }\href {https://doi.org/10.1038/s41563-025-02129-6} {10.1038/s41563-025-02129-6} (\bibinfo {year} {2025})\BibitemShut {NoStop}%
\bibitem [{\citenamefont {Lin}\ \emph {et~al.}(2024)\citenamefont {Lin}, \citenamefont {Sun}, \citenamefont {Dirnberger}, \citenamefont {Li}, \citenamefont {Qu}, \citenamefont {Wen}, \citenamefont {Sofer}, \citenamefont {Söll}, \citenamefont {Winnerl}, \citenamefont {Helm}, \citenamefont {Zhou}, \citenamefont {Dan},\ and\ \citenamefont {Prucnal}}]{lin_strong_2024}%
  \BibitemOpen
  \bibfield  {author} {\bibinfo {author} {\bibfnamefont {K.}~\bibnamefont {Lin}}, \bibinfo {author} {\bibfnamefont {X.}~\bibnamefont {Sun}}, \bibinfo {author} {\bibfnamefont {F.}~\bibnamefont {Dirnberger}}, \bibinfo {author} {\bibfnamefont {Y.}~\bibnamefont {Li}}, \bibinfo {author} {\bibfnamefont {J.}~\bibnamefont {Qu}}, \bibinfo {author} {\bibfnamefont {P.}~\bibnamefont {Wen}}, \bibinfo {author} {\bibfnamefont {Z.}~\bibnamefont {Sofer}}, \bibinfo {author} {\bibfnamefont {A.}~\bibnamefont {Söll}}, \bibinfo {author} {\bibfnamefont {S.}~\bibnamefont {Winnerl}}, \bibinfo {author} {\bibfnamefont {M.}~\bibnamefont {Helm}}, \bibinfo {author} {\bibfnamefont {S.}~\bibnamefont {Zhou}}, \bibinfo {author} {\bibfnamefont {Y.}~\bibnamefont {Dan}},\ and\ \bibinfo {author} {\bibfnamefont {S.}~\bibnamefont {Prucnal}},\ }\href {https://doi.org/10.1021/acsnano.3c07236} {\bibfield  {journal} {\bibinfo  {journal} {ACS Nano}\ }\textbf {\bibinfo {volume} {18}},\ \bibinfo {pages} {2898} (\bibinfo {year} {2024})}\BibitemShut
  {NoStop}%
\bibitem [{\citenamefont {Rozenman}\ \emph {et~al.}(2018)\citenamefont {Rozenman}, \citenamefont {Akulov}, \citenamefont {Golombek},\ and\ \citenamefont {Schwartz}}]{rozenman2018long}%
  \BibitemOpen
  \bibfield  {author} {\bibinfo {author} {\bibfnamefont {G.~G.}\ \bibnamefont {Rozenman}}, \bibinfo {author} {\bibfnamefont {K.}~\bibnamefont {Akulov}}, \bibinfo {author} {\bibfnamefont {A.}~\bibnamefont {Golombek}},\ and\ \bibinfo {author} {\bibfnamefont {T.}~\bibnamefont {Schwartz}},\ }\href@noop {} {\bibfield  {journal} {\bibinfo  {journal} {ACS photonics}\ }\textbf {\bibinfo {volume} {5}},\ \bibinfo {pages} {105} (\bibinfo {year} {2018})}\BibitemShut {NoStop}%
\bibitem [{\citenamefont {Hou}\ \emph {et~al.}(2020)\citenamefont {Hou}, \citenamefont {Khatoniar}, \citenamefont {Ding}, \citenamefont {Qu}, \citenamefont {Napolov}, \citenamefont {Menon},\ and\ \citenamefont {Forrest}}]{hou2020ultralong}%
  \BibitemOpen
  \bibfield  {author} {\bibinfo {author} {\bibfnamefont {S.}~\bibnamefont {Hou}}, \bibinfo {author} {\bibfnamefont {M.}~\bibnamefont {Khatoniar}}, \bibinfo {author} {\bibfnamefont {K.}~\bibnamefont {Ding}}, \bibinfo {author} {\bibfnamefont {Y.}~\bibnamefont {Qu}}, \bibinfo {author} {\bibfnamefont {A.}~\bibnamefont {Napolov}}, \bibinfo {author} {\bibfnamefont {V.~M.}\ \bibnamefont {Menon}},\ and\ \bibinfo {author} {\bibfnamefont {S.~R.}\ \bibnamefont {Forrest}},\ }\href@noop {} {\bibfield  {journal} {\bibinfo  {journal} {Advanced Materials}\ }\textbf {\bibinfo {volume} {32}},\ \bibinfo {pages} {2002127} (\bibinfo {year} {2020})}\BibitemShut {NoStop}%
\bibitem [{\citenamefont {Balasubrahmaniyam}\ \emph {et~al.}(2023)\citenamefont {Balasubrahmaniyam}, \citenamefont {Simkhovich}, \citenamefont {Golombek}, \citenamefont {Sandik}, \citenamefont {Ankonina},\ and\ \citenamefont {Schwartz}}]{balasubrahmaniyam2023enhanced}%
  \BibitemOpen
  \bibfield  {author} {\bibinfo {author} {\bibfnamefont {M.}~\bibnamefont {Balasubrahmaniyam}}, \bibinfo {author} {\bibfnamefont {A.}~\bibnamefont {Simkhovich}}, \bibinfo {author} {\bibfnamefont {A.}~\bibnamefont {Golombek}}, \bibinfo {author} {\bibfnamefont {G.}~\bibnamefont {Sandik}}, \bibinfo {author} {\bibfnamefont {G.}~\bibnamefont {Ankonina}},\ and\ \bibinfo {author} {\bibfnamefont {T.}~\bibnamefont {Schwartz}},\ }\href@noop {} {\bibfield  {journal} {\bibinfo  {journal} {Nature Materials}\ }\textbf {\bibinfo {volume} {22}},\ \bibinfo {pages} {338} (\bibinfo {year} {2023})}\BibitemShut {NoStop}%
\bibitem [{\citenamefont {Xu}\ \emph {et~al.}(2023)\citenamefont {Xu}, \citenamefont {Mandal}, \citenamefont {Baxter}, \citenamefont {Cheng}, \citenamefont {Lee}, \citenamefont {Su}, \citenamefont {Liu}, \citenamefont {Reichman},\ and\ \citenamefont {Delor}}]{xu2023ultrafast}%
  \BibitemOpen
  \bibfield  {author} {\bibinfo {author} {\bibfnamefont {D.}~\bibnamefont {Xu}}, \bibinfo {author} {\bibfnamefont {A.}~\bibnamefont {Mandal}}, \bibinfo {author} {\bibfnamefont {J.~M.}\ \bibnamefont {Baxter}}, \bibinfo {author} {\bibfnamefont {S.-W.}\ \bibnamefont {Cheng}}, \bibinfo {author} {\bibfnamefont {I.}~\bibnamefont {Lee}}, \bibinfo {author} {\bibfnamefont {H.}~\bibnamefont {Su}}, \bibinfo {author} {\bibfnamefont {S.}~\bibnamefont {Liu}}, \bibinfo {author} {\bibfnamefont {D.~R.}\ \bibnamefont {Reichman}},\ and\ \bibinfo {author} {\bibfnamefont {M.}~\bibnamefont {Delor}},\ }\href@noop {} {\bibfield  {journal} {\bibinfo  {journal} {Nature Communications}\ }\textbf {\bibinfo {volume} {14}},\ \bibinfo {pages} {3881} (\bibinfo {year} {2023})}\BibitemShut {NoStop}%
\bibitem [{\citenamefont {Steger}\ \emph {et~al.}(2013)\citenamefont {Steger}, \citenamefont {Liu}, \citenamefont {Nelsen}, \citenamefont {Gautham}, \citenamefont {Snoke}, \citenamefont {Balili}, \citenamefont {Pfeiffer},\ and\ \citenamefont {West}}]{steger2013long}%
  \BibitemOpen
  \bibfield  {author} {\bibinfo {author} {\bibfnamefont {M.}~\bibnamefont {Steger}}, \bibinfo {author} {\bibfnamefont {G.}~\bibnamefont {Liu}}, \bibinfo {author} {\bibfnamefont {B.}~\bibnamefont {Nelsen}}, \bibinfo {author} {\bibfnamefont {C.}~\bibnamefont {Gautham}}, \bibinfo {author} {\bibfnamefont {D.~W.}\ \bibnamefont {Snoke}}, \bibinfo {author} {\bibfnamefont {R.}~\bibnamefont {Balili}}, \bibinfo {author} {\bibfnamefont {L.}~\bibnamefont {Pfeiffer}},\ and\ \bibinfo {author} {\bibfnamefont {K.}~\bibnamefont {West}},\ }\href@noop {} {\bibfield  {journal} {\bibinfo  {journal} {Physical Review B}\ }\textbf {\bibinfo {volume} {88}},\ \bibinfo {pages} {235314} (\bibinfo {year} {2013})}\BibitemShut {NoStop}%
\bibitem [{\citenamefont {Zaitsev}\ \emph {et~al.}(2015)\citenamefont {Zaitsev}, \citenamefont {Il’ynskaya}, \citenamefont {Koudinov}, \citenamefont {Poletaev}, \citenamefont {Nikitina}, \citenamefont {Egorov}, \citenamefont {Kavokin},\ and\ \citenamefont {Seisyan}}]{zaitsev2015diffusive}%
  \BibitemOpen
  \bibfield  {author} {\bibinfo {author} {\bibfnamefont {D.}~\bibnamefont {Zaitsev}}, \bibinfo {author} {\bibfnamefont {N.}~\bibnamefont {Il’ynskaya}}, \bibinfo {author} {\bibfnamefont {A.}~\bibnamefont {Koudinov}}, \bibinfo {author} {\bibfnamefont {N.}~\bibnamefont {Poletaev}}, \bibinfo {author} {\bibfnamefont {E.}~\bibnamefont {Nikitina}}, \bibinfo {author} {\bibfnamefont {A.~Y.}\ \bibnamefont {Egorov}}, \bibinfo {author} {\bibfnamefont {A.}~\bibnamefont {Kavokin}},\ and\ \bibinfo {author} {\bibfnamefont {R.}~\bibnamefont {Seisyan}},\ }\href@noop {} {\bibfield  {journal} {\bibinfo  {journal} {Scientific Reports}\ }\textbf {\bibinfo {volume} {5}},\ \bibinfo {pages} {11474} (\bibinfo {year} {2015})}\BibitemShut {NoStop}%
\bibitem [{\citenamefont {Wurdack}\ \emph {et~al.}(2021)\citenamefont {Wurdack}, \citenamefont {Estrecho}, \citenamefont {Todd}, \citenamefont {Yun}, \citenamefont {Pieczarka}, \citenamefont {Earl}, \citenamefont {Davis}, \citenamefont {Schneider}, \citenamefont {Truscott},\ and\ \citenamefont {Ostrovskaya}}]{wurdack2021motional}%
  \BibitemOpen
  \bibfield  {author} {\bibinfo {author} {\bibfnamefont {M.}~\bibnamefont {Wurdack}}, \bibinfo {author} {\bibfnamefont {E.}~\bibnamefont {Estrecho}}, \bibinfo {author} {\bibfnamefont {S.}~\bibnamefont {Todd}}, \bibinfo {author} {\bibfnamefont {T.}~\bibnamefont {Yun}}, \bibinfo {author} {\bibfnamefont {M.}~\bibnamefont {Pieczarka}}, \bibinfo {author} {\bibfnamefont {S.~K.}\ \bibnamefont {Earl}}, \bibinfo {author} {\bibfnamefont {J.~A.}\ \bibnamefont {Davis}}, \bibinfo {author} {\bibfnamefont {C.}~\bibnamefont {Schneider}}, \bibinfo {author} {\bibfnamefont {A.}~\bibnamefont {Truscott}},\ and\ \bibinfo {author} {\bibfnamefont {E.}~\bibnamefont {Ostrovskaya}},\ }\href@noop {} {\bibfield  {journal} {\bibinfo  {journal} {Nature communications}\ }\textbf {\bibinfo {volume} {12}},\ \bibinfo {pages} {5366} (\bibinfo {year} {2021})}\BibitemShut {NoStop}%
\bibitem [{\citenamefont {Cho}\ \emph {et~al.}(2023)\citenamefont {Cho}, \citenamefont {Shin}, \citenamefont {Sung},\ and\ \citenamefont {Gong}}]{cho2023ultra}%
  \BibitemOpen
  \bibfield  {author} {\bibinfo {author} {\bibfnamefont {H.}~\bibnamefont {Cho}}, \bibinfo {author} {\bibfnamefont {D.-J.}\ \bibnamefont {Shin}}, \bibinfo {author} {\bibfnamefont {J.}~\bibnamefont {Sung}},\ and\ \bibinfo {author} {\bibfnamefont {S.-H.}\ \bibnamefont {Gong}},\ }\href@noop {} {\bibfield  {journal} {\bibinfo  {journal} {Nanophotonics}\ }\textbf {\bibinfo {volume} {12}},\ \bibinfo {pages} {2563} (\bibinfo {year} {2023})}\BibitemShut {NoStop}%
\bibitem [{\citenamefont {Zhang}\ \emph {et~al.}(2025)\citenamefont {Zhang}, \citenamefont {Nilforoushan}, \citenamefont {Weidgans}, \citenamefont {Hirschmann}, \citenamefont {Gronwald}, \citenamefont {Mosina}, \citenamefont {Sofer}, \citenamefont {Mooshammer}, \citenamefont {Dirnberger},\ and\ \citenamefont {Huber}}]{Zhang2025}%
  \BibitemOpen
  \bibfield  {author} {\bibinfo {author} {\bibfnamefont {H.}~\bibnamefont {Zhang}}, \bibinfo {author} {\bibfnamefont {N.}~\bibnamefont {Nilforoushan}}, \bibinfo {author} {\bibfnamefont {C.}~\bibnamefont {Weidgans}}, \bibinfo {author} {\bibfnamefont {J.}~\bibnamefont {Hirschmann}}, \bibinfo {author} {\bibfnamefont {I.}~\bibnamefont {Gronwald}}, \bibinfo {author} {\bibfnamefont {K.}~\bibnamefont {Mosina}}, \bibinfo {author} {\bibfnamefont {Z.}~\bibnamefont {Sofer}}, \bibinfo {author} {\bibfnamefont {F.}~\bibnamefont {Mooshammer}}, \bibinfo {author} {\bibfnamefont {F.}~\bibnamefont {Dirnberger}},\ and\ \bibinfo {author} {\bibfnamefont {R.}~\bibnamefont {Huber}},\ }\href {https://doi.org/10.48550/arXiv.2506.06010} {\bibinfo {title} {Exciton-polariton condensates in van der {{Waals}} magnetic microwires}} (\bibinfo {year} {2025}),\ \Eprint {https://arxiv.org/abs/2506.06010} {arXiv:2506.06010 [cond-mat]} \BibitemShut {NoStop}%
\end{thebibliography}%

\clearpage

% Figures tables and captions
% Permission statements are required for all figures reproduced or adapted from previously published articles/sources. Please also ensure that all necessary permissions to reproduce images have been received
% Please remove these statements for original figures

% % Please provide Biographies and photos for Essays, Feature Articles, Progress Reports, Reviews, and Perspectives for those authors who should be highlighted  
% % These should be at most 100 words long
% % For other article types this section can be removed
% % Photographs should be 40mm broad and 50 mm high

% \begin{figure}
%   \includegraphics{bio-placeholder.jpg}
%   \caption*{Biography}
% \end{figure}

% \begin{figure}
%   \includegraphics{bio-placeholder.jpg}
%   \caption*{Biography}
% \end{figure}

% \begin{figure}
%   \includegraphics{bio-placeholder.jpg}
%   \caption*{Biography}
% \end{figure}

% \begin{figure}
%   \includegraphics{bio-placeholder.jpg}
%   \caption*{Biography}
% \end{figure}

% Table of contents entry should be 50 - 60 words long
% Image should be 55 mm broad and 50 mm high or 110 mm broad and 20 mm high

%\begin{figure}
%\textbf{Table of Contents}\\
%\medskip
%  \includegraphics{toc-image.png}
%  \medskip
%  \caption*{ToC Entry}
%\end{figure}

\end{document}

% -------- Page 1: Title --------
\section*{Supporting Information}

\vspace{1em}

\noindent\textbf{Directional Flow of Confined Polaritons in CrSBr}

\vspace{1em}

\noindent\textit{Pratap Chandra Adak, Sichao Yu, Jaime Abad-Arredondo, Biswajit Datta, Andy Cruz, Sorah Fischer, Kseniia Mosina, Zdeněk Sofer, Antonio I. Fernández-Domínguez, Francisco J. Garc\'ia-Vidal, Vinod M. Menon}

\vspace{2em}

\newpage

% -------- Page 2: Table of Contents --------
\tableofcontents
\newpage

% -------- Page 3+: Main Content --------
\section{Setup for the optical measurements}
\begin{figure}[hbt!]
    \centering
	\includegraphics[width=15cm]{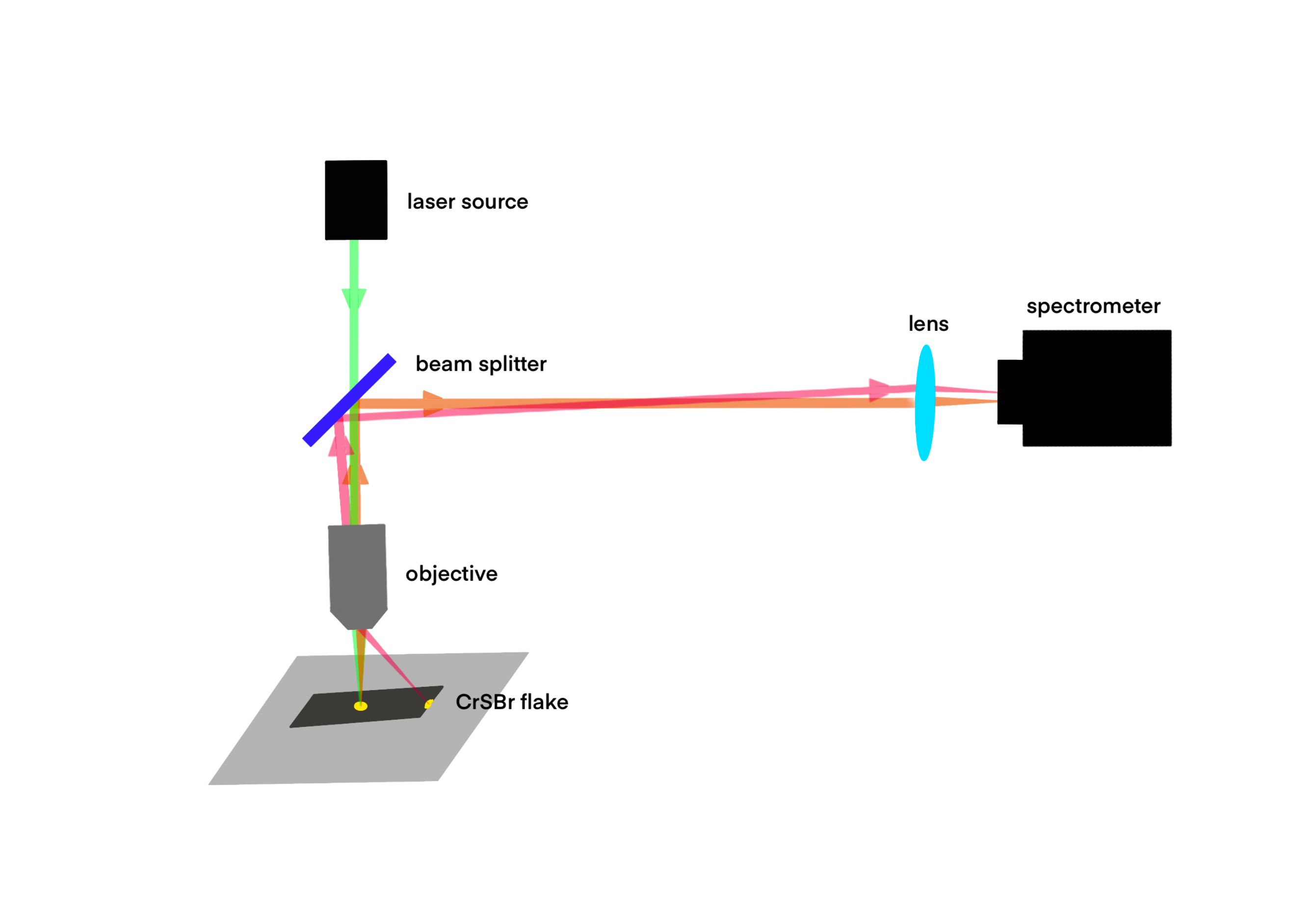}
        {
        \caption{Simplified schematic of the setup for propagation measurements.  
        }
        \label{SIfig:fig Propagation setup}
        }
\end{figure}

Figure~\ref{SIfig:fig Propagation setup} shows the setup of our propagation measurements. The CrSBr sample was excited by the laser (green line) and PL signal from the excitation point (orange line) and edge of the flake (red line) were collected simultaneously by the spectrometer.

\begin{figure}[hbt!]
    \centering
	\includegraphics[width=15cm]{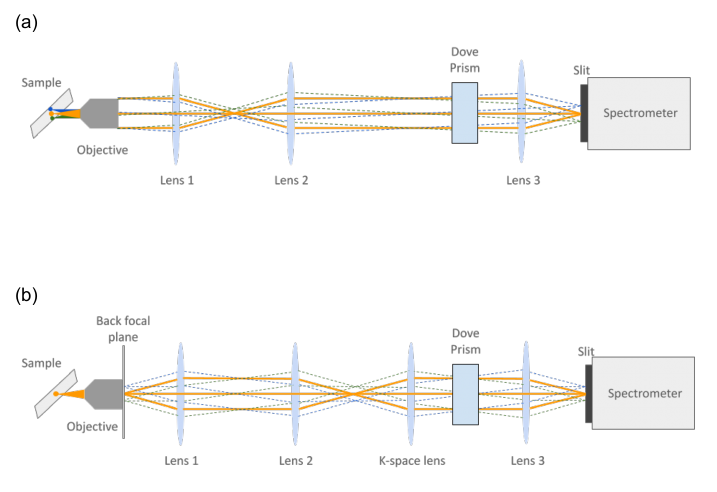}
        {
        \caption{Simplified schematic of the setup for spatially-resolved measurements (a) and momentum-resolved measurements (b)
        }
        \label{SIfig:fig Confinement setup}
        }
\end{figure}

For spatially resolved measurements, PL emissions from different positions on the sample are focused onto different regions of the spectrometer, corresponding to their spatial origin. The slit in front of the spectrometer helps select PL signals along a specific direction of the flake, depending on the orientation between the flake and the slit. Figure~\ref{SIfig:fig Confinement setup}a shows the configuration in which the slit is oriented along the \textit{b}-axis of the CrSBr crystal, allowing only PL signals from points along the \textit{b}-axis to pass through the slit. The selected signals are then analyzed by the spectrometer, revealing the spatial distribution of the PL spectrum along the chosen direction.

To measure the momentum-resolved spectrum, an additional lens is required to image the back focal plane of the objective onto the spectrometer (Figure~\ref{SIfig:fig Confinement setup}b). The slit helps select signals along a specific direction on the back focal plane, revealing the angular distribution of emission along a corresponding momentum direction.

The dove prism is placed before the Lens 3 to rotate the light beam to change the orientation between the slit and sample (spatially resolved) or the back focal plane (momentum-resolved).

\begin{figure}[hbt!]
    \centering
	\includegraphics[width=8cm]{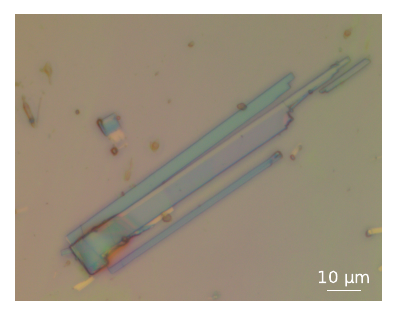}
        {
        \caption{Microscopic image of CrSBr flakes for the polariton confinement measurements
        }
        \label{SIfig:fig Confinement sample}
        }
\end{figure}

Figure~\ref{SIfig:fig Confinement sample} shows the CrSBr flakes used to fabricate the DBR/CrSBr/DBR microcavities for confinement measurements. The top flake is 5 µm wide, and the bottom flake is 2.6 µm wide.

\section{Additional PL propagation image}

\begin{figure}[hbt!]
    \centering
	\includegraphics[width=10cm]{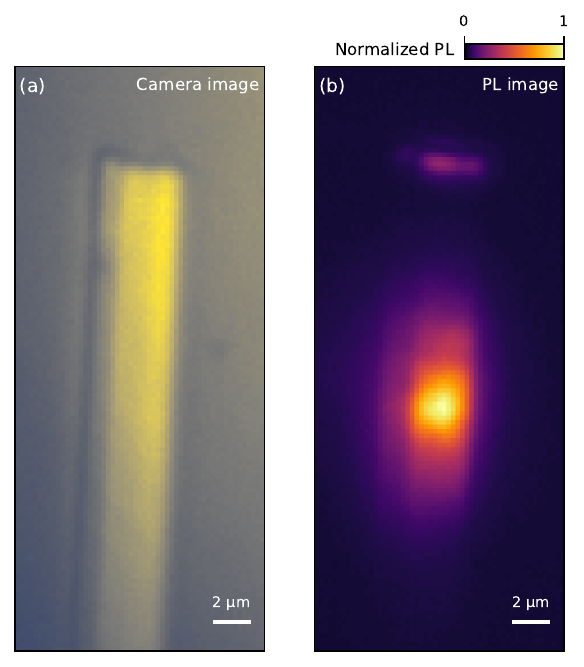}
        {
        \caption{Camera image (a) and PL image (b) show the polariton propagate along a strip-shaped CrSBr flake
        }
        \label{SIfig:fig Propagation image}
        }
\end{figure}

\begin{figure}[hbt!]
    \centering
	\includegraphics[width=10cm]{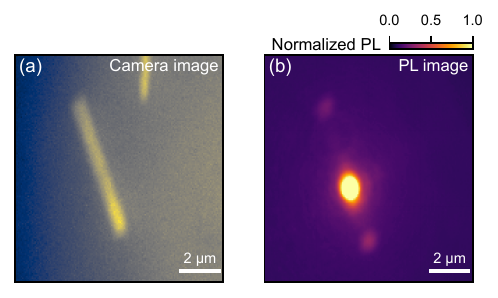}
        {
        \caption{Camera image (a) and PL image (b) for another CrSBr flakes showing the polariton propagation
        }
        \label{SIfig:fig More propagation image 01}
        }
\end{figure}

Polariton propagation is a general phenomenon observed in CrSBr flakes. Figure~\ref{SIfig:fig Propagation image} shows the PL image collected from a 100~nm thick flake, demonstrating polariton propagation along the \textit{a}-axis. This is also the same flake that we used in the main text to measure polariton propagation over different distances. Figure~\ref{SIfig:fig More propagation image 01} offers additional evidence supporting the presence of PL propagation in CrSBr flakes.

\section{Polariton branches identification}

\begin{figure}[hbt!]
    \centering
	\includegraphics[width=15cm]{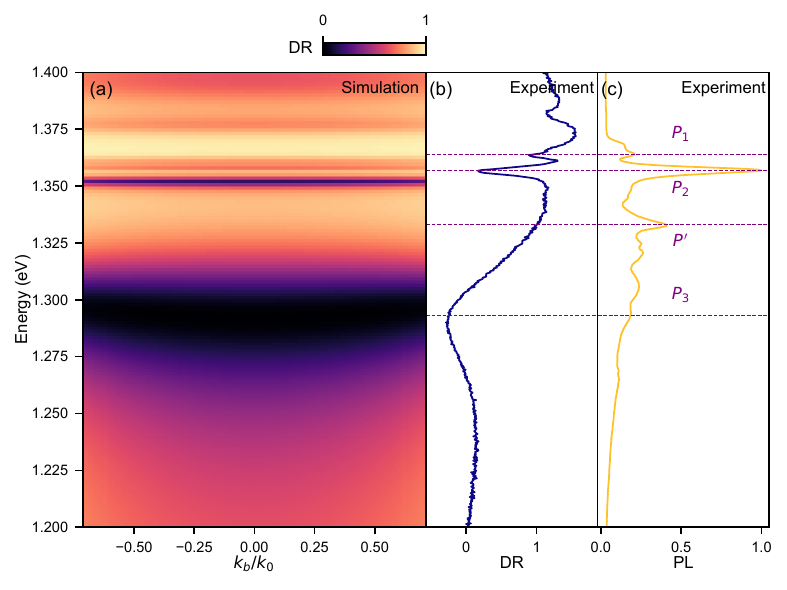}
        {
        \caption{
        (\textbf{a})~Simulated momentum-resolved reflectivity on a 115 nm-thick CrSBr flake. The simulation was performed using transfer-matrix method.
        (\textbf{b, c})~Reflectivity (b) and PL spectrum (c) measured from a 115 nm-thick CrSBr flake. 
        }
        \label{SIfig:fig RATcatcher simulation}
        }
\end{figure}

To identify the polariton branches in the sample that we discuss in the main manuscript, we examine both the reflectivity and the PL spectrum from a flake with similar thickness.
Figure~\ref{SIfig:fig RATcatcher simulation}b,c show the results from this flake with a thickness of 115~nm.
For the $P_1$, $P_2$, and $P_3$ peaks mentioned in the main text, both the reflectivity and PL spectra clearly exhibit these features, confirming their identification as polariton branches. In contrast, the $P'$ peak observed in the PL spectrum does not appear in the reflectivity data, suggesting that it may originate from a defect-related state or a surface exciton.
Transfer-matrix based simulation using MATLAB  (Figure~\ref{SIfig:fig RATcatcher simulation}a) further supports this interpretation: despite some discrepancies between the simulation and experimental data, it clearly shows that only $P_1$, $P_2$, and $P_3$ correspond to polariton branches.

\section{Details of propagation length calculations}

\begin{figure}[hbt!]
    \centering
	\includegraphics[width=15cm]{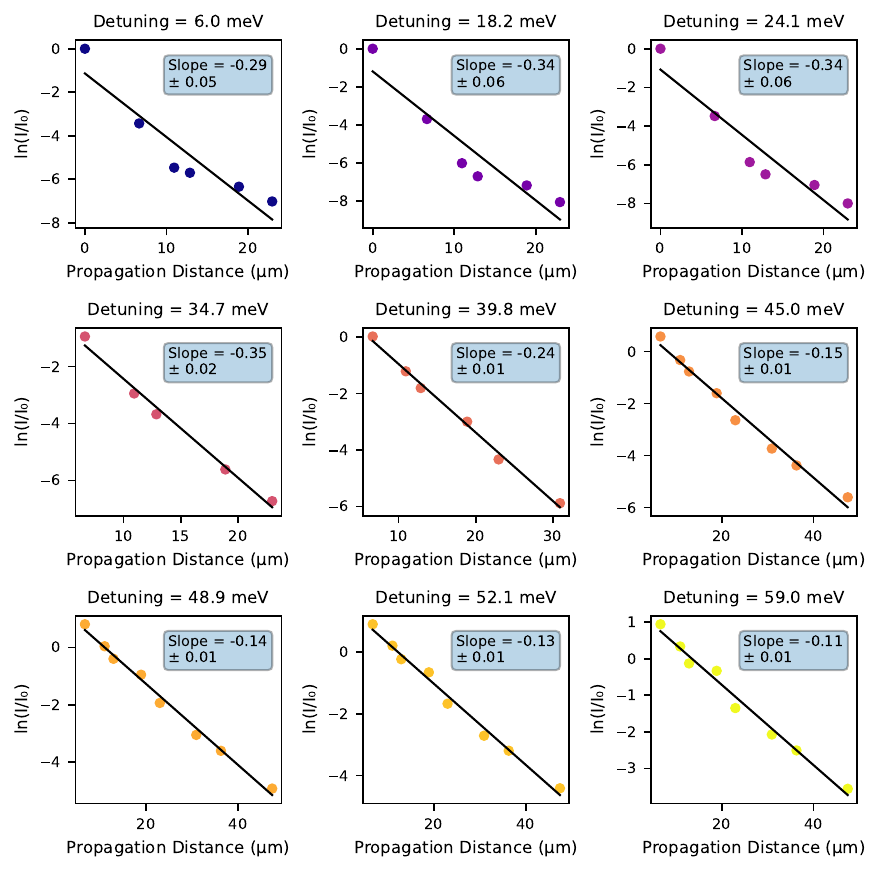}
        {
        \caption{Logarithmic plot of the normalized photoluminescence (PL) intensity after propagation as a function of distance. The vertical axis represents \(\ln(I/I_0)\), where \(I\) is the PL intensity after propagation and \(I_0\) is the intensity at the corresponding excitation spot. A linear fit is applied to extract the the propagation length of the polariton branch.
        }
        \label{SIfig:fig Fitting results}
        }
\end{figure}

Since the PL intensity decays as \( R \propto e^{-x/L_\text{prop}} \), we plotted the logarithm of the normalized PL intensity as a function of propagation distance and applied a linear fit to extract the propagation length \( L_\text{prop} \) as the inverse of the slope. Figure~\ref{SIfig:fig Fitting results} shows the details of the fit. For polariton branches with large exciton components, the PL intensity decays rapidly, and the linear fit yields large errors, indicating that these polaritons are more localized. In contrast, the photon-like polaritons exhibit larger propagation length, revealing their intrinsically higher propagation efficiency.

\clearpage

\clearpage

\section{Theoretical modeling of propagation length}

The propagation length and mode profiles presented in Figure 2 of the main text are obtained using numerical mode analysis and analytical dispersion calculations as outlined below.
\subsection{Numerical simulations}
To model the guided modes of the flake, we use the Mode Analysis simulation in \textit{COMSOL Multiphysics}. The simulation parameters and geometry are specified as follows:
\begin{itemize}
    \item The flake cross section is modeled as a rectangle with height $H = 100\,\mathrm{nm}$ and width $W = 6.3\,\mu\mathrm{m}$.
    \item This cross section lies in the $b$--$c$ plane of the crystal, with guided modes propagating along the $a$-axis with a phase factor of the form $\exp(i k_0 n_{\mathrm{eff}} x)$, where $n_{\mathrm{eff}}$ is the effective index of the mode, and $k_0 = 2\pi / \lambda$ is the free-space wavevector.
    \item At each wavelength of interest, the simulation identifies all guided modes supported by the flake structure by obtaining their associated $n_{\mathrm{eff}}$ and mode profile.
\end{itemize}
We then numerically determine the propagation length of a mode. Since fields propagate as $e^{i k_0 n_{\mathrm{eff}} x}$, then the intensity profile evolves as $e^{- 2 k_0 \mathrm{Im}(n_{\mathrm{eff}}) x}$, and thus the propagation length is given simply by:
\begin{gather}
L_{\mathrm{prop}}=\frac{\lambda_0}{4\pi\,\mathrm{Im}(n_{\mathrm{eff}})},
\end{gather}
with $\lambda_0$ being the free-space wavelength of light.

\subsection{Analytical dispersion for TE modes in an infinite Slab}
To validate and complement the numerical results, we also consider the analytical dispersion relation for TE-polarized modes supported by an infinite anisotropic dielectric slab embedded in an uniform background medium. This represents the limit of infinite width of the flakes studied in the numerical simulations. The guided modes propagate along the $a$-axis with a phase factor of the form $\exp(i k_0 n_{\mathrm{eff}} x)$, where $n_{\mathrm{eff}}$ is the effective index of the mode, and $k_0 = 2\pi / \lambda$ is the free-space wavevector. The dispersion relation for the TE modes (field along the b direction of the crystal) is given by:
\begin{equation}
\tan\left(n_{\mathrm{bg}} k_0 H \sqrt{\frac{\varepsilon_b}{\varepsilon_{\mathrm{bg}}} - \left(\frac{n_{\mathrm{eff}}}{n_{\mathrm{bg}}}\right)^2}\right)
= 
\frac{2 \sqrt{ \left(\left(\frac{n_{\mathrm{eff}}}{n_{\mathrm{bg}}}\right)^2 - 1\right) \left(\frac{\varepsilon_b}{\varepsilon_{\mathrm{bg}}} - \left(\frac{n_{\mathrm{eff}}}{n_{\mathrm{bg}}}\right)^2 \right) }}{1 + \frac{\varepsilon_b}{\varepsilon_{\mathrm{bg}}} - 2 \left(\frac{n_{\mathrm{eff}}}{n_{\mathrm{bg}}}\right)^2},
\end{equation}
where in this expression $\varepsilon_b$ is the permittivity of the slab material (along the $b$-axis),$\varepsilon_{\mathrm{bg}}$ is the permittivity of the surrounding background medium and $n_{\mathrm{bg}} = \sqrt{\varepsilon_{\mathrm{bg}}}$. This transcendental equation is solved numerically using a custom implementation of the Delves-Lyness method\textsuperscript{~\cite{Delves1967}}, which efficiently locates the complex roots of the analytic function corresponding to $n_{\mathrm{eff}}$.

\section{Confinement and photoluminescence simulations}

\begin{figure}[hbt!]
    \centering
	\includegraphics[width=15cm]{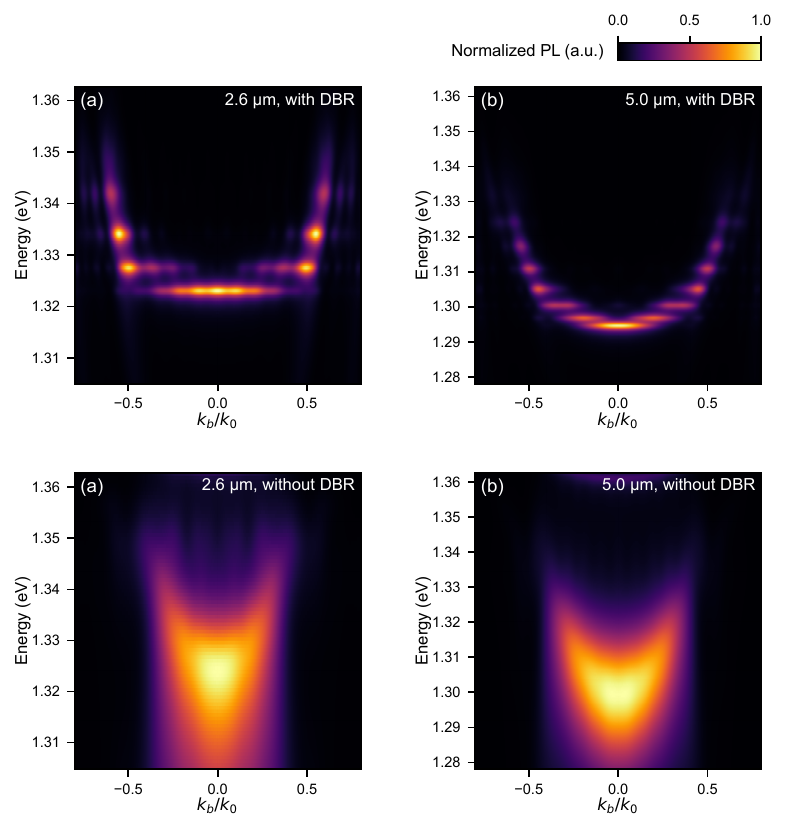}
        {
        \caption{Simulated momentum-resolved PL spectra data for the 2.6 µm-wide (a, c) and 5 µm-wide (b, d) CrSBr with and without top DBR. Momutum \(k\) oriented along the \(k_b\) of the Fourier space.
        }
        \label{SIfig:fig Simulation no DBR}
        }
\end{figure}

To study the effects of vertical confinement on the photoluminescence (PL) of CrSBr flakes, we perform frequency-domain electromagnetic simulations using \textit{COMSOL Multiphysics}. We model the system under the assumption of translational invariance along the $a$-axis, allowing us to reduce the problem to a two-dimensional (2D) geometry in the $b$--$c$ plane. All simulations follow a common procedure described below, and are performed in two distinct structural configurations.

In the first configuration, the CrSBr flake is deposited directly on a substrate composed of a 90 nm thick silica (SiO$_2$) layer atop a silicon (Si) base. This structure supports slab resonances due to the dielectric discontinuities at the interfaces. Nevertheless, the spatial mode structure of these resonances cannot be resolved from far-field measurements alone.

In the second configuration, the CrSBr flake is vertically confined within a photonic cavity formed by distributed Bragg reflector (DBR) mirrors. The bottom DBR is composed of 11 alternating pairs of SiO$_2$ and TiO$_2$ layers, while the top DBR contains 5 alternating pairs of SiO$_2$ and SiN. The stopband central wavelengths are centered at 935~nm for the bottom mirror and 930~nm for the top mirror. This entire DBR-encapsulated stack is placed on a silicon substrate.

In both configurations, PL emission is simulated by introducing an electric point dipole source inside the CrSBr slab. The dipole moment is oriented along the $b$-axis, consistent with the excitonic transition dipole of the material. To emulate spatially incoherent emission, we scan the dipole position laterally across the flake. For each dipole location, we compute the angular distribution of the far-field radiation using a projection technique. Since PL is inherently incoherent, the total emission profile is obtained by summing the far-field intensity contributions from all dipole positions. In Figure~\ref{SIfig:fig Simulation no DBR} we show the momentum-resolved simulated PL spectrum for both cases of DBR-encapsulated (top) and not encapsulated CrSBr flake, demonstrating how the extra vertical confinement from the DBR lead to the arising of clear resonances. 

This approach allows us to systematically probe the emission characteristics of modes in both the open and confined structures, and to directly compare how DBR confinement modifies the angular emission of the flake, finding good agreement with the experimental measurements.

\clearpage

\section{Additional confinement data on CrSBr microcavity}

\begin{figure}[hbt!]
    \centering
	\includegraphics[width=15cm]{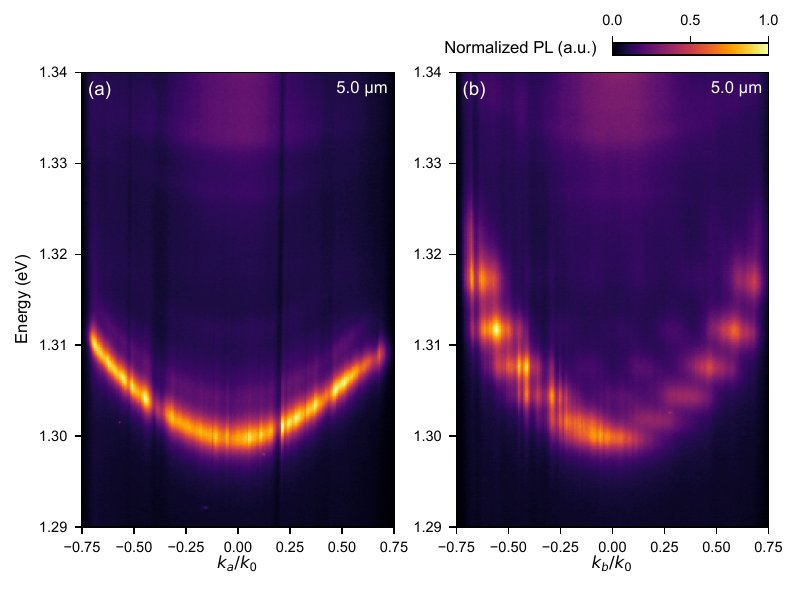}
        {
        \caption{Momentum-resolved PL spectra for the 5 µm-wide at the energy range from 1.29 eV to 1.34 eV. Momutum \(k\) oriented along the \(k_a\) (a) and \(k_b\) (b) of the Fourier space. 
        }
        \label{SIfig:fig Confinement full spectrum}
        }
\end{figure}

Figure~\ref{SIfig:fig Confinement full spectrum} shows momentum-resolved spectroscopy from another spot on the 5 µm-wide flake in the energy range from 1.29 eV to 1.34 eV. 
As discussed in the main manuscript, we see a prominent continuous $E-k_a$ dispersion in Figure~\ref{SIfig:fig Confinement full spectrum}a. Figure~\ref{SIfig:fig Confinement full spectrum}b shows the corresponding $E-k_b$ dispersion with energy quantized at discrete $k_b$.
These discrete $k_b$ modes on Figure~\ref{SIfig:fig Confinement full spectrum}b have their counterpart in Figure~\ref{SIfig:fig Confinement full spectrum}a showing as weak parabolic dispersions, parallel to the prominent lowest-energy mode.

\begin{figure}[hbt!]
    \centering
	\includegraphics[width=12cm]{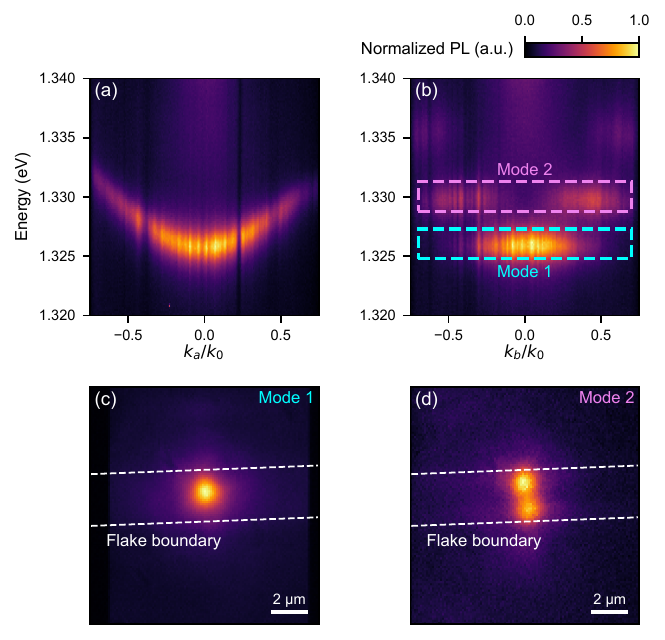}
        {
        \caption{
        (\textbf{a,b})~Momentum-resolved PL spectra were measured on a 2.6 µm-wide CrSBr flake inside DBR-DBR cavity, with momentum \(k\) oriented along the \(k_a\) (a) and \(k_b\) (b) of the Fourier space. 
        (\textbf{c,d})~Real-space PL distribution maps of the 2.6\,\textmu m-wide CrSBr flake, corresponding to mode 1 (c) and mode 2 (d) highlighted in panel (b).
        }
        \label{SIfig:fig Confinement second flake}
        }
\end{figure}

Figure~\ref{SIfig:fig Confinement second flake} shows the momentum-resolved PL spectra measured on a 2.6 µm-wide CrSBr flake. Compared to the 5\,\textmu m-wide flake discussed in the main text, discrete polariton modes, at different energies and fewer in number, are observed.
These differences from the 5~\textmu m flake indicate that the flake width serves as an effective parameter for tuning polariton confinement in CrSBr. To further investigate the confined modes in this flake, we employed a combination of a long-pass and a short-pass filter to isolate emissions within specific energy ranges and measure the real space image of the PL emission. The resulting PL distribution maps for mode~1 (c) and mode~2 (d) correspond to the boxed regions highlighted in Figure~\ref{SIfig:fig Confinement second flake}b. 

\begin{figure}[hbt!]
    \centering
	\includegraphics[width=15cm]{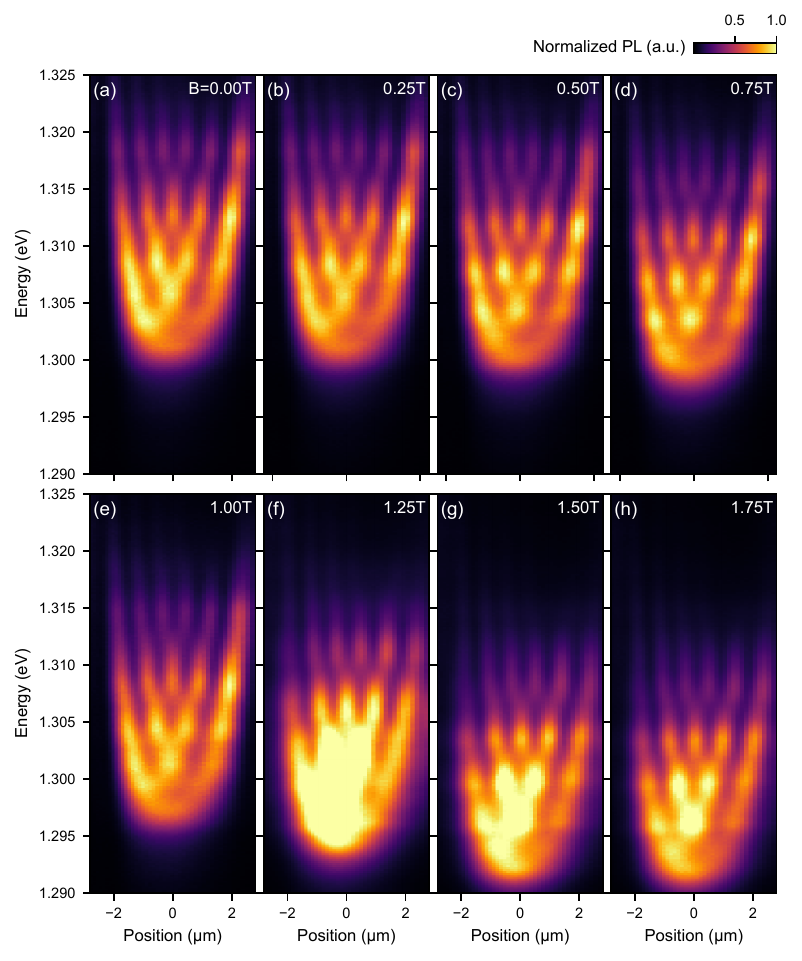}
        {
        \caption{Experimental measured spatially resolved PL spectra of a 5.0 µm-wide flake with different magnetic field. The magnetic field applied along the \textit{c}-axis of the CrSBr crystal. The applied magnetic field \( B \) was varied from 0 T to 1.75 T in increments of 0.25 T. The spectrum exhibits an approximately 10 meV redshift under the applied magnetic field.
        }
        \label{SIfig:fig Real space magnetic}
        }
\end{figure}

Figure~\ref{SIfig:fig Real space magnetic} shows how the spatially-resolved PL spectra measured on the 5 µm-wide flake respond to the applied magnetic field. The overall spectrum redshifts as the magnetic field increases, showing behavior consistent with the momentum-resolved measurements and reinforcing the conclusion that this type of confinement can be tuned by the magnetic field.

\begin{figure}[hbt!]
    \centering
	\includegraphics[width=15cm]{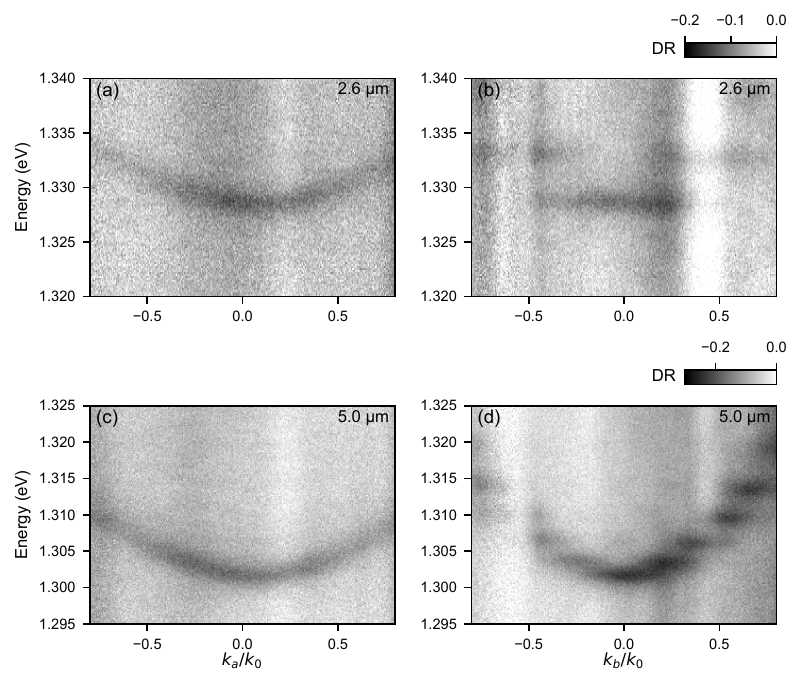}
        {
        \caption{Momentum-resolved reflectivity measured on the 2.6 µm-wide flake (a, b) and 5 µm-wide flake (c, d). The light used for the measurement is from a broadband CW white light source with with a linear polarizer to align the input light polarization along the \textit{b}-axis of the CrSBr. For panels (a) and (c), the spectrometer slit is aligned along the \textit{a}-axis of the flakes, whereas for panels (b) and (d), it is aligned along the \textit{b}-axis.
        }
        \label{SIfig:fig Confinement reflectivity}
        }
\end{figure}

Figure~\ref{SIfig:fig Confinement reflectivity} shows the white-light reflectivity data measured on the 2.6 µm-wide and 5 µm-wide flakes, with the slit oriented along both the \textit{a}-axis and \textit{b}-axis of each flake. We use a CW white-light source, with a linear polarizer to align the input light polarization along the \textit{b}-axis. The reflectivity data from the sample were normalized to \( R_0 \), the reflectivity measured from the empty cavity. The reflectivity data exhibit the same pattern as the PL spectrum, providing further evidence for the observed confinement.

\clearpage
\section{Refractive index change due to magnetic field}
\begin{figure}[hbt!]
    \centering
	\includegraphics[width=15cm]{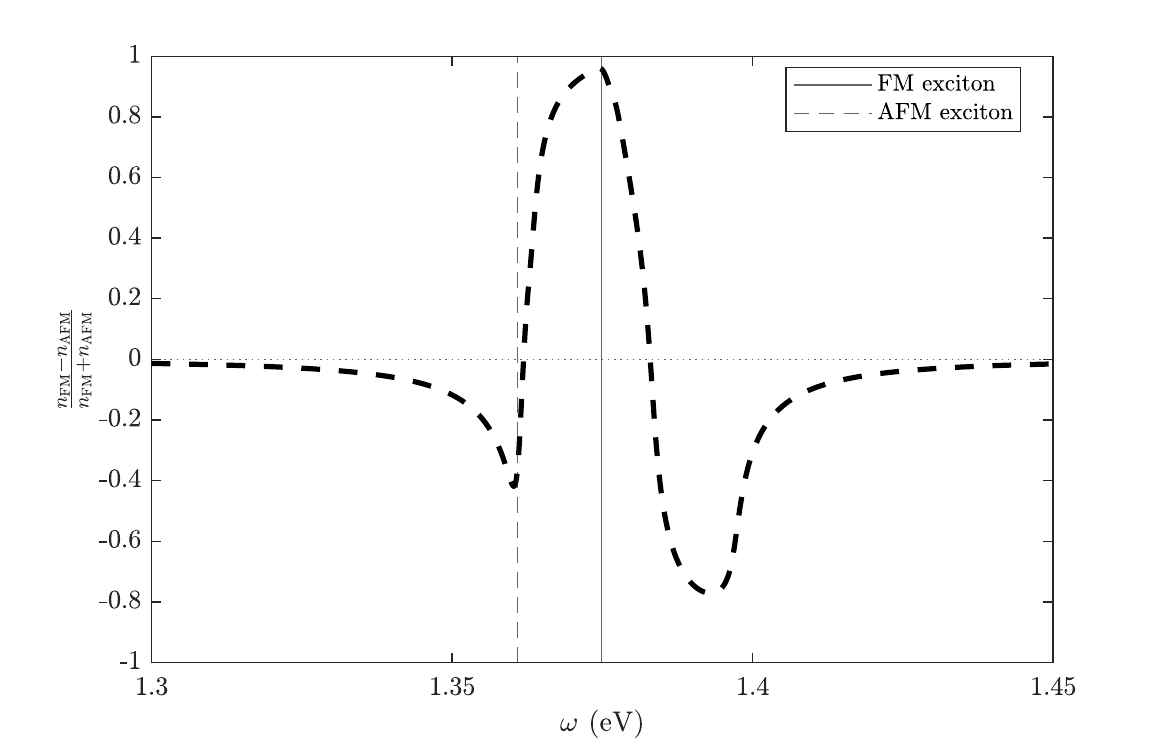}
        {
        \caption{The refractive index change of the material as it shifts from AFM to FM state.
        }
        \label{SIfig:fig Normalized_RI_change}
        }
\end{figure}

Upon the transition from antiferromagnetic to ferromagnetic order, the exciton energies redshift. This results in a remarkable change in the refractive index, as shown in Figure~\ref{SIfig:fig Normalized_RI_change}.

\clearpage
\bibliographystyle{MSP}
\bibliography{references}